\documentclass[Journal]{IEEEtran}
\pdfoutput=1
\usepackage{multirow}
\usepackage{booktabs}
\usepackage{amsfonts}
\usepackage{amsmath}
\usepackage{amsmath,cases}
\usepackage{amssymb}
\usepackage{graphicx}
\usepackage{cite}
\usepackage{epsfig}
\usepackage{url}
\usepackage{algorithm}
\usepackage{algorithmicx, algpseudocode}
\usepackage{epstopdf}
\usepackage{color}
\usepackage[tight]{subfigure}
\usepackage{mathrsfs}
\usepackage[justification=centering]{caption}
\usepackage{float}
\usepackage{diagbox}

\newcommand{\clA}{{\cal A}}

\newcommand{\ds}{\displaystyle}

\newcommand{\la}{\langle}
\newcommand{\ra}{\rangle}

\newcommand{\bx}{\boldsymbol{x}}

\newcommand{\clC}{{\cal C}}

\newcommand{\clN}{{\cal N}}

\newcommand{\clK}{{\cal K}}

\newcommand{\clU}{{\cal U}}

\newcommand{\bt}{\pmb{t}}

\newcommand{\diag}{{\sf diag}}

\newcommand{\bv}{\mathbf{v}}

\newcommand{\bp}{\mathbf{p}}

\newcommand{\by}{\mathbf{y}}

\newcommand{\clH}{{\cal H}}
\newcommand{\bP}{\mathbf{P}}
\newcommand{\bY}{\mathbf{Y}}

\newcommand{\xk}{x^{(\kappa)}}
\newcommand{\xko}{x^{(\kappa+1)}}

\newcommand{\bV}{\mathbf{V}}

\newcommand{\rk}{r^{(\kappa)}}
\newcommand{\ak}{a^{(\kappa)}}
\newcommand{\Bk}{B^{(\kappa)}}
\newcommand{\Ck}{C^{(\kappa)}}

\newcommand{\pk}{p^{(\kappa)}}
\newcommand{\pko}{p^{(\kappa+1)}}

\newcommand{\rhok}{\rho^{(\kappa)}}

\newcommand{\tk}{t^{(\kappa)}}

\newcommand{\btheta}{\pmb{\theta}}
\newcommand{\thetak}{\theta^{(\kappa)}}
\newcommand{\thetako}{\theta^{(\kappa+1)}}
\newcommand{\gammak}{\gamma^{(\kappa)}}

\newcommand{\ttak}{\tilde{\tilde{a}}^{(\kappa)}}

\newcommand{\ttbk}{\tilde{\tilde{b}}^{(\kappa)}}

\newcommand{\ttck}{\tilde{\tilde{c}}^{(\kappa)}}

\newcommand{\bk}{b^{(\kappa)}}

\newcommand{\gk}{g^{(\kappa)}}

\newcommand{\bbC}{\mathbb{C}}

\newcommand{\clAk}{\clA^{(\kappa)}}

\newcommand{\clCk}{\clC^{(\kappa)}}
\newcommand{\Psik}{\Psi^{(\kappa)}}
\newcommand{\phik}{\phi^{(\kappa)}}
\newcommand{\tpi}{\tilde{\pi}}
\newcommand{\tthetako}{\tilde{\theta}^{(\kappa+1)}}
\newcommand{\tthetak}{\tilde{\theta}^{(\kappa)}}
\newcommand{\psik}{\psi^{(\kappa)}}
\newcommand{\clE}{\mathcal{E}}
\newcommand{\tko}{t^{(\kappa+1)}}
\newcommand{\paik}{\pi^{(\kappa)}}
\newcommand{\gammako}{\gamma^{(\kappa+1)}}
\allowdisplaybreaks
\begin{document}
\title{RIS-aided Zero-Forcing and Regularized Zero-Forcing Beamforming in Integrated Information and Energy Delivery\thanks{This work was supported in part  by the Australian Research Councils Discovery Projects under Grant DP190102501,  and in part by the Engineering and Physical Sciences Research Council projects EP/P034284/1 and EP/P003990/1 (COALESCE), and European Research Council's Advanced Fellow Grant QuantCom (Grant No. 789028)}}
\author{H. Yu$^1$,  H. D. Tuan$^1$, E. Dutkiewicz$^1$, H. V. Poor$^2$, and L. Hanzo$^3$
\thanks{$^1$School of Electrical and Data Engineering, University of Technology Sydney, Broadway, NSW 2007, Australia (email:Hongwen.Yu@student.uts.edu.au, Tuan.Hoang@uts.edu.au, Eryk.Dutkiewicz@uts.edu.au); $^2$Department of Electrical and Computer Engineering, Princeton University, Princeton, NJ 08544, USA (email: poor@princeton.edu); $^3$School of Electronics and Computer Science, University of Southampton, Southampton, SO17 1BJ, UK (email: lh@ecs.soton.ac.uk)}
}
\date{}
\maketitle
\vspace*{-1cm}
\begin{abstract}
This paper considers a network of a multi-antenna array base station (BS) and
a reconfigurable intelligent  surface (RIS) to deliver both information to
information users (IUs) and power to  energy users (EUs). The RIS links the connection
between the IUs and the BS as there is no direct path between the former and the latter.
The EUs are located nearby the BS  in order to effectively harvest energy from the
high-power signal from the BS, while the much weaker signal reflected from the RIS hardly contributes
to the EUs' harvested energy.
To provide reliable links for all users over the
same time-slot, we adopt the transmit time-switching (transmit-TS) approach, under which information and energy are
delivered over different time-slot fractions. This allows us to rely on  conjugate beamforming  for
energy links and zero-forcing/regularized zero-forcing beamforming (ZFB/RZFB) and on the programmable reflecting coefficients (PRCs) of the RIS for information links. We show that  ZFB/RZFB and PRCs can be still separately optimized in their joint design, where PRC optimization is based on iterative closed-form expressions.
We then develop
a path-following algorithm for solving our
max-min IU throughput optimization problem subject to  a realistic constraint on the
quality-of-energy-service  in terms of the EUs' harvested energy thresholds. We also propose a new RZFB for substantially  improving the IUs' throughput.
\end{abstract}
\section{Introduction}
Jointly supporting both wireless information and power transfer networking poses challenging problems in  signal processing for communication (see e.g. \cite{Hu-18-FQ-A} and references therein). For information and energy delivery over a single time slot, simultaneous wireless information and power transfer (SWIPT) apportions the power of the  signals received by the users
for energy-harvesting (EH) and information detection (ID). In the context, the EH performance is dependent on the power of the received signal, by contrast, the ID performance is  critically dependent on the signal-to-interference-plus-noise ratio (SINR). The popular SWIPT systems have primarily used conjugate beamforming (CB) to deliver sufficient energy for EH \cite{Zhao-16-Feb-A}, even though this limits the ID performance due to the multi-user interference (MUI) imposed. By contrast, zero-forcing beamforming (ZFB) completely eliminates the MUI, but it is less efficient for SWIPT. To circumvent these drawbacks, it has been proposed to convey information and energy over the same time slot by transmitting information and energy in separate fractions of the time-slot. Termed as transmit time-switching (transmit-TS), it has been shown to outperform SWIPT due to its ability to support individual energy beamforming for EH and  information beamforming for ID (see e.g.  \cite{Naetal17,Taetal17a,Yuetal20tcom,Yuetal20tvt}). As a benefit, transmit-TS enables CB for EH and ZFB or regularized zero-forcing beamforming (RZFB) for ID \cite{PHS05,NTDP19}\cite{Yuetal20tvt}.

A reconfigurable intelligent surface (RIS) is a planar array of  "nearly-passive" reflecting elements, which can beneficially manipulate the reflected signals  by programming its reflection coefficients \cite{TAD17}.  By strategically installing a RIS in  places
such as building facades so that it can have a line-of-sight (LoS) path from both the users and a base station (BS), the RIS facilitates reliable communication when
there are no direct links between them \cite{Huetal18,WZ19Mag,Reetal19,Huaetal21}. A challenging problem in signal processing for these RIS-aided networks is to jointly design the BS's transmit beamformers and the RIS's programmable reflecting coefficients (PRCs) to maximizing the sum throughput \cite{Huetal18,Nadetal20,Zhoetal20a,Panetal20} or the users' minimum throughput \cite{Yuetal21}. The joint design of power allocation for ZFB and PRCs to maximize the sum throughput subject to  individual user throughput constraints has been considered in \cite{Huetal18}. While the alternating  power allocation optimization with the PRCs held fixed is simple,
the alternating optimization in PRCs with the power allocation held fixed is very challenging since
the user throughput becomes a complex function due to the matrix inversion involved in ZFB. As a result, the convergence behavior of the general purpose gradient descent algorithm used in \cite{Huetal18} is unknown. Thus, the expected computational tractability of the ZFB design could not be achieved.
The authors of \cite{Paetal19power} and \cite{WZ20jsac} considered some RIS-aided SWIPT scenarios which
require that  the BS, the RIS and the energy users (EUs) must be located within a small  cell radius of about $10$m, however  the reflected signal by the RIS after undergoing the associated double path-loss becomes quite weak compared to that coming directly from the BS to the EUs, which erodes the benefit of the RIS in EH.

\begin{figure}[!htb]
\centering
\includegraphics[width=8cm]{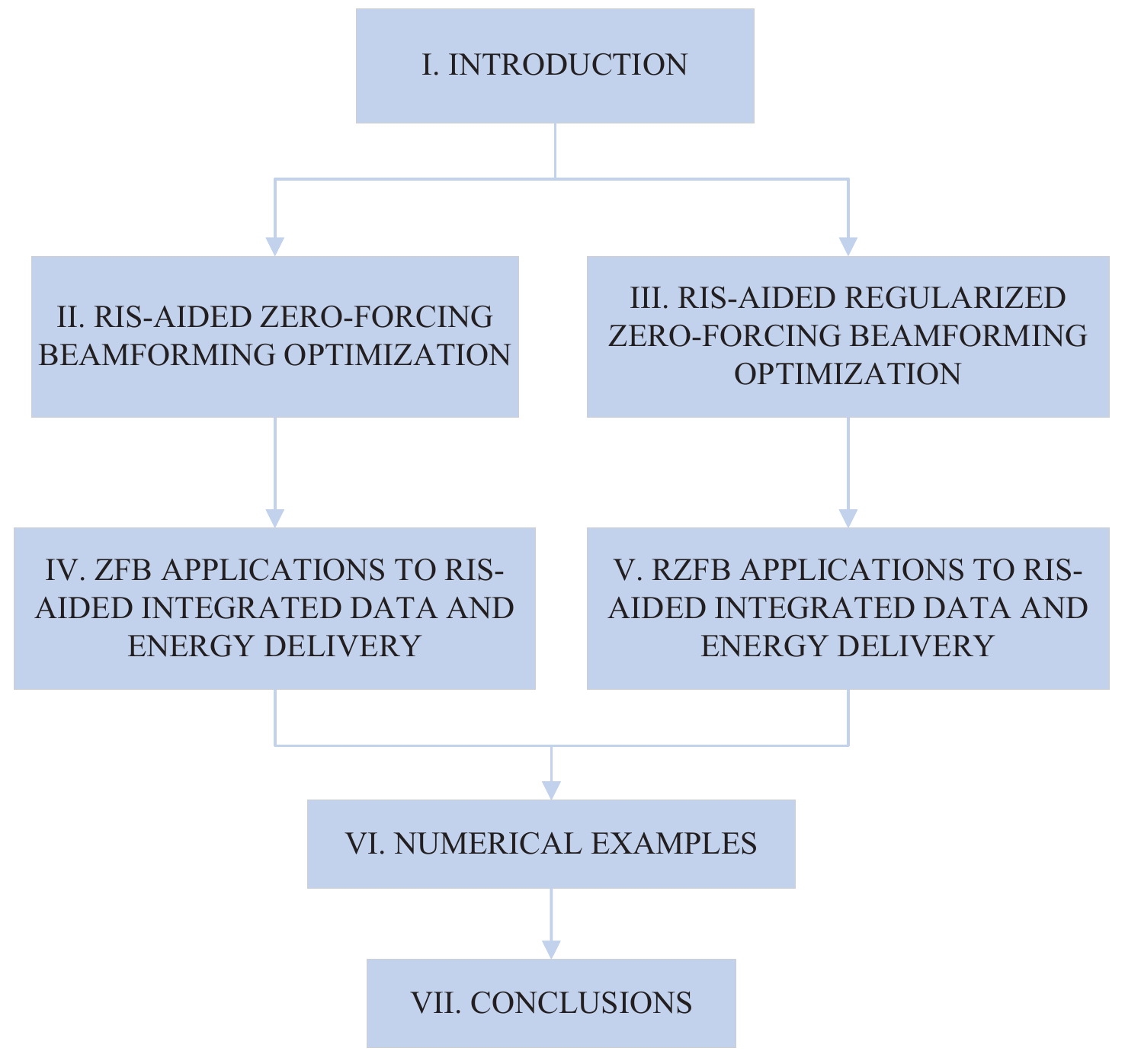}
\caption{Paper organization.}\label{outline}
\end{figure}

Against the above background, this paper offers the following contributions.
\begin{itemize}
\item We  reveal that as a benefit of the transmit-TS technique, the joint optimization of  power allocations (for ZFB) and PRCs may be simplified to optimizing the PRCs only, because the power allocation for ZFB can be easily determined. Instead of iterating by relying on convex problems or using deep Q-learning methods to handle
the unit-modulus constraint on the PRCs which incur much higher computational complexity \cite{chongwen1,Yuetal21,chongwen6},  we use the polar form of unit-modulus complex numbers that allows each descent iteration of the RIS coefficient calculation to be based on computational efficient closed-form expressions for the solution of concave trigonometric function optimization \cite{Tuybook}. Accordingly, we develop efficient computational procedures, which are based on closed-form expressions for its computation;

\item Similarly, we also show that the joint optimization of power allocations (for RZFB) and PRCs can
be decomposed into the separate optimization of power allocations (for RZFB) and optimization of PRCs. Accordingly, we develop efficient computational procedures for PRC optimization, which are still based on closed-form expressions. A computational procedure is also proposed for power allocations optimization, which involves a convex quadratic problem at each iteration;
\item Furthermore, we develop a new RZFB for improving the throughput of IUs. Our simulations show that the IUs' throughput using the new RZFB  is 15\% - 25\%  higher than that obtained by the existing RZFB in the challenging rank-deficient scenario, when the BS only has a few antennas for serving more IUs;
\item We consider a practical scenario of RIS-aided  integrated information and energy deliveries to  both
information users (IUs) and energy users (EUs). By adopting the aforementioned transmit-TS approach, we harness CB for delivering energy to EUs, and RIS-aided ZFB/RZFB/new RZFB for delivering information to IUs.
Naturally, the PRCs are still separately optimized in this joint design problem.  We then develop efficient computational procedures for solving the problem of maximizing the IUs' minimum throughput subject  to a constraint on the  quality-of-energy-service (QoES) in terms of the EUs' harvested energy thresholds.
\end{itemize}
The rest of the paper is organized as follows. Section II and Section III are respectively
devoted to PRC optimization for ZFB and RZFB with  its applications to RIS-aided information and energy delivery
studied in Section IV and V. A new RZFB is also introduced in Section V. Section VI provides simulations to support
the technical developments of the previous sections. Section VII concludes the paper. The Appendix provides several inequalities that are frequently used in the technical sections.  The  flow chart of paper organization can be seen in Fig. \ref{outline}.

{\it Notation.} Only design variables are denoted in bold face; ${\cal C}(0,a)$ is the set of circular Gaussian random variables with  zero means and variances $a$; $\angle x_i$ is the argument of the complex number $x_i$
and as such $\angle x$ for $x=(x_1,\dots, x_N)^T\in\mathbb{C}^N$ is understood componentwise:  $\angle x\triangleq (\angle x_1,\dots,\angle x_N)^T\in\mathbb{C}^{N}$; $I_N$ is the identity matrix of size $N\times N$, while $O_{M\times N}$ is a zero matrix of size $M\times N$; for $x=(x_1,\dots, x_N)^T$, ${\sf diag}(x)$ is a diagonal matrix of  size $N\times N$ with $x_1, x_2, \dots, x_N$ on its diagonal; we also write $\la X\ra=\mbox{trace}(X)$ for notational simplicity; $[X]^2$ is $XX^H$, so $[X^H]^2=X^HX$, and $\la X,Y\ra=\la X^HY\ra$ for  matrices $X$ and $Y$; accordingly, the Frobenius norm of $X$ is defined by $||X||=\sqrt{\la[X]^2\ra}$; the notation $X \succeq 0$ ($X\succ 0$, resp.) used for the Hermitian symmetric matrix $X$ indicates that it is positive definite (positive semi-definite, resp.); the maximal eigenvalue of the Hermitian symmetric matrix $X$ is denoted by $\lambda_{\max}(X)$.
\section{RIS-aided zero-forcing beamforming optimization}
\begin{figure}[!htb]
\centering
\includegraphics[width=8cm]{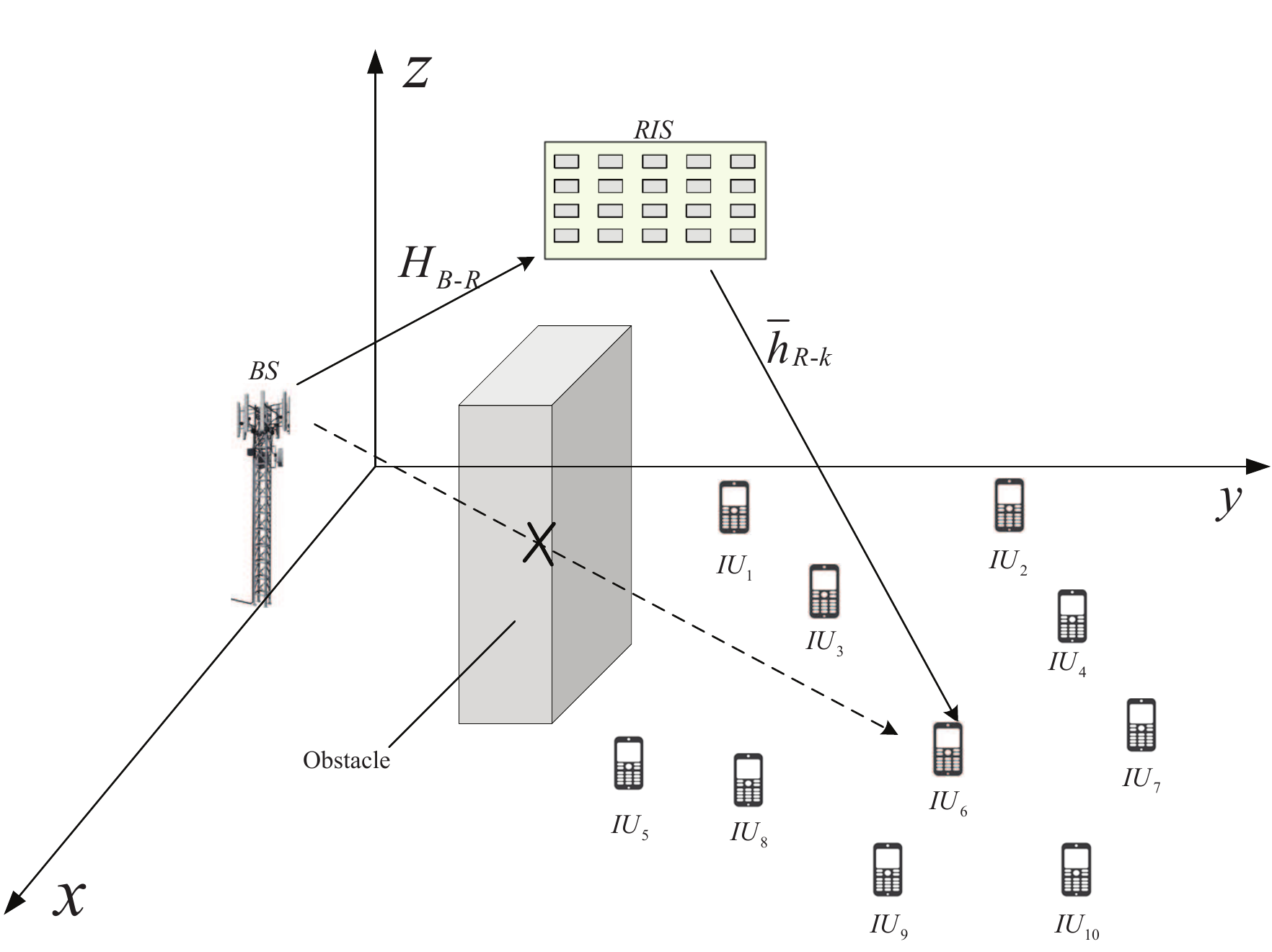}
\caption{Scenario setup with the blockage of the direct path between the BS and the IUs.}\label{s2}
\end{figure}

Consider a RIS-aided network, which is illustrated by Fig. \ref{s2}
with a RIS of  $N$ reflecting units to assist the downlink  from an $M$-antenna base station (BS) to $K$ single-antenna information users (IUs) $k\in\clK \triangleq \{1, \dots, K\}$ because there is no direct signal path between the former and the  latter.\footnote{According to \cite{Yuetal21,Yutwc}, the networks throughput is hardly improved by RIS's, when there are direct paths from the BS to the IUs.}
The channel spanning from the BS
to the RIS is $\tilde{H}_{B-R}\triangleq \sqrt{\beta_{B-R}}H_{B-R}\in\mathbb{C}^{N\times M}$, where $\sqrt{\beta_{B-R}}$ models the path-loss and large-scale fading of LoS and the entries of $H_{B-R}$ are $\clC(0,1)$, modelling small-scale fading. Analogously,
the channel spanning from the RIS
to IU $k$ is $\tilde{h}_{R-k}=\sqrt{\beta_{R-k}}\bar{h}_{R-k}\in \mathbb{C}^{1\times N}$, where
$\sqrt{\beta_{R-k}}$  represents the large-scale fading,  while
$\bar{h}_{R-k}$  denotes the small-scale fading having elements of ${\cal C}(0,1)$. Like in many other papers on RIS-aided communication networks, we assume perfect channel state information, which can be obtained by channel estimation \cite{chongwen02,WZ20jsac,chongwen1}.

Let $\btheta\triangleq (\btheta_1,\dots, \btheta_N)\in [0,2\pi]^N$ and
\[
e^{\jmath\btheta}\triangleq (e^{\jmath\btheta_1},\dots,  e^{\jmath\btheta_N})^T\in\mathbb{C}^{N},
\]
which is a complex-vector function of the variable $\btheta$. We also define the diagonal complex-matrix function of the variable $e^{\jmath\btheta}$ as
\[
\diag[e^{\jmath\btheta}]=\diag\{e^{\jmath \btheta_1},\dots, e^{\jmath \btheta_N}\}\in\bbC^{N\times N}
\]
representing the matrix of RIS reflection-coefficients.

The received signal at IU $k$ is
\begin{eqnarray}
y_k&=&h_{B-k}(e^{\jmath\btheta})x_I+n_k,\label{lb3}
\end{eqnarray}
where $n_k\in\clC(0,\sigma)$ is the background noise, and
\begin{equation}\label{lb4}
h_{B-k}(\btheta)\triangleq \tilde{h}_{R-k}R_{RIS-k}^{1/2}\diag[e^{\jmath\btheta}]\tilde{H}_{B-R}\in\mathbb{C}^{1\times M},
\end{equation}
which is the composite channel spanning from the BS to IU $k$.

In (\ref{lb4}), $R_{RIS-k} \in \bbC^{N\times N}$ represents the spatial correlation matrix of the RIS elements with respect to IU $k$ \cite{Nadetal20}.
For
\[
y\triangleq \begin{bmatrix}y_1\cr
\dots\cr
y_K\end{bmatrix}\in\mathbb{C}^{K},
\bar{n}\triangleq \begin{bmatrix}n_1\cr
\dots\cr
n_K\end{bmatrix}\in\mathbb{C}^{K},
\]
and
\begin{eqnarray}
\clH(e^{\jmath\btheta})&\triangleq& \begin{bmatrix}h_{B-1}(e^{\jmath\btheta})\cr
\dots\cr
h_{B-K}(e^{\jmath\btheta})\end{bmatrix}=H_R\diag[e^{\jmath\btheta}]\tilde{H}_{B-R} \nonumber \\
&=&\sum_{n=1}^Ne^{\jmath\theta_n}\clH_n
\in\mathbb{C}^{K\times M} \nonumber
\end{eqnarray}
with
\[
\clH_n\triangleq H_R\Psi_n\tilde{H}_{B-R}\in\mathbb{C}^{K\times M}\],
\[
H_R\triangleq \begin{bmatrix}\tilde{h}_{R-1}R_{RIS-1}^{1/2}\cr
\dots\cr
\tilde{h}_{R-K}R_{RIS-K}^{1/2}
\end{bmatrix}\in \mathbb{C}^{K\times N},
\]
where  $\Psi_n$ is a matrix of size $N\times N$ with all-zero entries, excepts $\Psi_n(n,n)=1$.

We can write
\begin{eqnarray}\label{lb5}
y=\clH(e^{\jmath\btheta})x_I+\bar{n}.
\end{eqnarray}
Now, for $K\leq M$ we consider the ZFB, under which the BS transmits
\begin{eqnarray}\label{zf1}
x_I&=&\clH^H(e^{\jmath\btheta})\left([\clH(e^{\jmath\btheta})]^2\right)^{-1}\mbox{diag}[\bp_k]_{k=1,\dots, K}s,
\end{eqnarray}
where  $s=(s_1,\dots, s_K)^T$, $s_k\in \mathcal{C}(0,1)$ is the information intended for the IUs having
the power of $\bp=(\bp_1,\dots, \bp_K)^T$. Then Equation (\ref{lb5}) becomes
\begin{eqnarray}
y&=&\clH(e^{\jmath\btheta})\clH^H(e^{\jmath\btheta})\left([\clH(e^{\jmath\btheta})]^2\right)^{-1}\mbox{diag}[\bp_k]_{k=1,\dots, K}s+\bar{n} \nonumber\\
&=&[\clH(e^{\jmath\btheta})]^2\left([\clH(e^{\jmath\btheta})]^2\right)^{-1}\mbox{diag}[\bp_k]_{k=1,\dots, K}s+\bar{n} \nonumber\\
&=&\mbox{diag}[\bp_k]_{k=1,\dots, K}s+\bar{n}\label{lb5e}
\end{eqnarray}
simplifying (\ref{lb3})  to
\begin{eqnarray}
y_k&=&\bp_ks_k+\bar{n}_k.\label{lb8}
\end{eqnarray}
The throughput for $s_k$ is
\begin{equation}\label{p1}
\ln(1+\bp_k^2/\sigma),
\end{equation}
and the transmit power is
\begin{eqnarray}
\mathbb{E}(||x_I||^2)&=&\la \clH^H(e^{\jmath\btheta})\left([\clH(e^{\jmath\btheta})]^2\right)^{-1}\mbox{diag}[\bp^2_k]_{k=1,\dots, K} \nonumber\\
&&\left([\clH(e^{\jmath\btheta})]^2\right)^{-1}\clH(e^{\jmath\btheta})\ra\label{p2e} \\
&=&\la \mbox{diag}[\bp^2_k]_{k=1,\dots, K}\left([\clH(e^{\jmath\btheta})]^2\right)^{-1}\ra.\label{p2}
\end{eqnarray}
For the IUs' max-min throughput optimization is employed, which aims for maximizing the users' worst-case (minimal) throughput,  where we have $\bp_k\equiv \bp_0$
\footnote{From (\ref{p1}), the users' worst-case throughput
is $\min_{k=1,\dots, K}\ln(1+\bp_k^2/\sigma)$, which
is maximized at $\ln(1+\bp_1^2/\sigma)= \ln(1+\bp_2^2/\sigma)=\dots=\ln(1+\bp_K^2/\sigma)$
$\Leftrightarrow$  $\bp_k\equiv \bp_0$}($\ln(1+\bp^2_k/\sigma)\equiv \ln(1+\bp^2_0/\sigma)$. Then by (\ref{p2}), the transmit power is $\bp_0^2\la \left([\clH(e^{\jmath\btheta})]^2\right)^{-1}\ra$, and the problem of max-min IU throughput optimization subject to a transmit power budget $P$ can be formulated as
\begin{eqnarray}
&&\max_{\bp_0,\btheta}\ \ln(1+\bp_0^2/\sigma) \quad\mbox{s.t.}\quad
\bp_0^2\la \left([\clH(e^{\jmath\btheta})]^2\right)^{-1}\ra\leq P\qquad\label{p3}\\
&&\Leftrightarrow\max_{\btheta}\ P/\la \left([\clH(e^{\jmath\btheta})]^2\right)^{-1}\ra\label{p3e}\\
&&\Leftrightarrow\min_{\btheta} f(e^{\jmath\btheta})\triangleq \la  \left(\clH(e^{\jmath\btheta})\clH^H(e^{\jmath\btheta})\right)^{-1}\ra.\label{p4}
\end{eqnarray}
In fact, it follows from the power constraint in (\ref{p3}) that $\bp_0^2\leq P/\la \left([\clH(e^{\jmath\btheta})]^2\right)^{-1}\ra$. Hence the problem (\ref{p3}) is actually
$\ds\max \ln(1+P/(\sigma\la \left([\clH(e^{\jmath\btheta})]^2\right)^{-1}\ra))$, which is the same as (\ref{p3e}). Since only the denominator of the fractional objective function in (\ref{p3e}) is dependent on $\btheta$, its maximization is equivalent to the minimization of its denominator, which is (\ref{p4}).

The rest of this section is devoted to the detailed portrayal of our algorithms conceived for computing (\ref{p4}), which is very challenging
because its objective function is highly nonlinear and computationally intractable.
\subsection{Step descent algorithm}
Let $\thetak$ be a point found during the $(\kappa-1)$-st iteration. The linearized function of $f$ at
$e^{\jmath\thetak}$ is
\begin{eqnarray}
3f(e^{\jmath\thetak})-2\Re\{\la [\clH^H(e^{\jmath\thetak})\left([\clH(e^{\jmath\thetak})]^2\right)^{-2}\clH(e^{\jmath\btheta}) \ra\}&=&\nonumber\\
3f(e^{\jmath\thetak})-2\sum_{n=1}^N\Re\{e^{\jmath\theta_n}\la \clH^H(e^{\jmath\thetak})\clAk\clH_n\ra  \},&&
\label{p6}
\end{eqnarray}
for
\begin{equation}\label{p7}
\clAk\triangleq \left([\clH(e^{\jmath\thetak})]^{2}\right)^{\color{black}-2}.
\end{equation}
We seek a step descent by addressing the following problem
\begin{equation}\label{p9}
\max_{\btheta}\ \sum_{n=1}^N\Re\{e^{\jmath\theta_n}\la \clH^H(e^{\jmath\thetak})\clAk\clH_n\ra  \},
\end{equation}
 which is decomposed into $N$ independent problems:
\[
\max_{\btheta_n}\ \Re\{e^{\jmath\theta_n}\la \clH^H(e^{\jmath\thetak})\clAk\clH_n\ra  \}, n=1,\dots, N,
\]
each of which admits the closed-form solution
\begin{equation}\label{p10}
\tthetako_n=-\angle \la \clH^H(e^{\jmath\thetak})\clAk\clH_n  \ra, n=1,\dots, N.
\end{equation}
We may then choose $\thetako$ according to one of the following rules:
\begin{itemize}
\item The simplest one
\begin{equation}\label{spl}
\thetako=\tthetako.
\end{equation}
\item Considering $\psik\triangleq \tthetako-\thetak$ as a step descent, we update $\thetako$ according to the so-called Barzilai-Borwein (BB) step size of \cite{BB88} in (\ref{gr1s}) and (\ref{gr1}).
\begin{figure*}[t]
\begin{eqnarray}
\thetako&=&\thetak +\frac{|\la \psik,\psik-\psi^{(\kappa-1)}\ra|}{||\psik-\psi^{(\kappa-1)}||^2}
\psik\label{gr1s}\\
&=&\thetak +\frac{|\la \tthetako-\thetak, \tthetako-\thetak{\color{black}-}\tthetak{\color{black}+}\theta^{(\kappa-1)}\ra|}{||\tthetako-\thetak-
\tthetak+\theta^{(\kappa-1)}||^2}(\tthetako-\thetak).\label{gr1}
\end{eqnarray}
\end{figure*}
\item Considering $\psik\triangleq e^{\jmath\tthetako}-e^{\jmath\thetak}$ we update $\thetako$ according to (\ref{gr1a}) and (\ref{gr1b}).We will refer this as the projective Barzilai-Borwein (PBB) step size.
\begin{figure*}[t]
\begin{eqnarray}
\thetako&=&\angle\left(e^{\jmath\thetak} +\frac{|\la \psik,\psik-\psi^{(\kappa-1)}\ra|}{||\psik-\psi^{(\kappa-1)}||^2}
\psik   \right)\label{gr1a}\\
&=&\angle\left(e^{\jmath\thetak} +\frac{|\la e^{\jmath\tthetako}-e^{\jmath\thetak}, e^{\jmath\tthetako}-e^{\jmath\tthetak}-e^{\jmath\tthetak}+e^{\jmath\theta^{(\kappa-1)}}\ra|}
{||e^{\jmath\tthetako}-e^{\jmath\tthetak}-e^{\jmath\tthetak}+e^{\jmath\theta^{(\kappa-1)}}||^2}\right.\left.(e^{\jmath\tthetako}-e^{\jmath\thetak})      \right).\label{gr1b}
\end{eqnarray}
\hrulefill
\end{figure*}

\end{itemize}
Algorithm \ref{alg1} provides the pseudo-code for the procedure iterating (\ref{spl}) or (\ref{gr1}), or (\ref{gr1b}) in order to arrive at the computational solution of (\ref{p4}). The reader is referred to \cite{BB88} for the rationale behind them in locating better feasible points, which are suitable for unconstrained optimization only.
There is an explicit update of the incumbent point in Algorithm \ref{alg1} because the updating rules (\ref{spl})-(\ref{gr1b}) do not enhance that $\thetak$ is the incumbent. Somewhat surprisingly, the performance of
Algorithm \ref{alg1} was found to be indifferent with using any of three aforementioned rules.
\begin{algorithm}
	\caption{ZFB step descent algorithm for (\ref{p4})} \label{alg1}
	\begin{algorithmic}[1]
	\State \textbf{Initialization:} Initial $\theta^{(0)}$ and set $\theta^{opt}=\theta^{(0)}$ and $\gamma^{opt}=
f(\theta^{opt})$ as the incumbent RIS and value.
		\State \textbf{Repeat until convergence of $\thetak$:} Generate $\tthetako$ by (\ref{p10}). Then generate
$\thetako$ either by (\ref{spl}) or (\ref{gr1}), or (\ref{gr1b}).
If $f(\thetako)<\gamma^{opt}$, set $\theta^{opt}=\thetako$ and $\gamma^{opt}=f(\thetako)$.
 Set $\kappa:=\kappa+1$.
 \State \textbf{Output} $\theta^{opt}$ and $\gamma^{opt}$.
	\end{algorithmic}
\end{algorithm}
\subsection{Full step descent algorithms}
We express $f$ in (\ref{p4}) as:
\begin{eqnarray}
f(e^{\jmath\btheta})&=&\alpha ||e^{\jmath\btheta}||^2  -\left(\alpha ||e^{\jmath\btheta}||^2 - \la\left([\clH(e^{\jmath\btheta})]^2\right)^{-1}\ra\right)\quad\label{n1}\\
&=&\alpha N -g(e^{\jmath\btheta}),\label{n1e}
\end{eqnarray}
where $\alpha>0$ is chosen for ensuring that the function
\[
g(e^{\jmath\btheta})\triangleq \alpha ||e^{\jmath\btheta}||^2 - \la\left([\clH(e^{\jmath\btheta})]^2\right)^{-1}\ra
\]
is convex in $e^{\jmath\btheta}$. The problem (\ref{p4}) is equivalent to the following problem
of unconstrained concave optimization \cite{Tuybook}\footnote{(\ref{n2}) is equivalent to $\min_{\btheta} (-g(e^{\jmath\btheta})$,
where $-g(e^{\jmath\btheta})$ is a concave function}
\begin{equation}\label{n2}
\max_{\btheta}\ g(e^{\jmath\btheta}).
\end{equation}
Following \cite{ATcdc98,AT,AT00} we will develop a local Frank-and-Wolf (FW) feasible direction algorithm for solving this problem as  it bypasses the line search to give as a full step size of length $1$. Moreover, this kind of FW
algorithm has proved to be very efficient in practice \cite{BM93}. To this end, let
$\thetak$ be a point found during the $(\kappa-1)$-st iteration.
Note that as $g$ is convex, its linearized function provides its lower bound formulated as:
\begin{eqnarray}
&&g(e^{\jmath\btheta}) \nonumber \\&\geq& \alpha\left(2\Re\{\sum_{n=1}^Ne^{\jmath\theta_n}e^{-\jmath\thetak_n}\}-N\right) \nonumber -
3\la\left([\clH(e^{\jmath\thetak})]^2\right)^{-1}\ra\nonumber\\
&&+2\Re\{\sum_{n=1}^Ne^{\jmath\theta_n}\la \clH^H(e^{\jmath\thetak})\clAk\clH_n\ra  \}\label{n3}\\
&=&-\alpha N-3\la\left([\clH(e^{\jmath\thetak})]^2\right)^{-1}\ra \nonumber \\&&+2\sum_{n=1}^N\Re\{e^{\jmath\theta_n}\left(\alpha e^{-\jmath\thetak_n}+\la \clH^H(e^{\jmath\thetak})\clAk\clH_n\ra\right)    \}\nonumber\\
&\triangleq&\gk(e^{\jmath\btheta}),\label{n3a}
\end{eqnarray}
where $\clAk$ is defined in (\ref{p7}). For finding the FW feasible direction, we
solve the following problem at the $\kappa$-th iteration to generate $\thetako$
\begin{equation}\label{n4}
\max_{\btheta}\ 2\sum_{n=1}^N\Re\{e^{\jmath\theta_n}\left(\alpha e^{-\jmath\thetak_n}+\la \clH^H(e^{\jmath\thetak})\clAk\clH_n\ra\right)    \},
\end{equation}
 which admits the following closed-form solution similar to (\ref{p10}):
\begin{eqnarray}\label{n5}
\thetako_n=-\angle\left( \alpha e^{-\jmath\thetak_n}+\la \clH^H(e^{\jmath\thetak})\clAk\clH_n\ra  \right),\nonumber \\ n=1,\dots, N.
\end{eqnarray}
We can readily show that
\begin{equation}\label{n5a}
g(e^{\jmath\thetako})\geq \gk(e^{\jmath\thetako})>\gk(e^{\jmath\thetak})=g(e^{\jmath\thetak}),
\end{equation}
so $\thetako$ is a better point than $\thetak$, i.e. $1=\mbox{arg}\max_{0\leq\pmb{\nu}\leq 1}
g(e^{\jmath(\thetak+\pmb{\nu}(\thetako-\thetak))})$, so the full step size of length one is achieved.
\footnote{Obviously, the step size is not full whenever $\mbox{arg}\max_{0\leq\pmb{\nu}\leq 1}
g(e^{\jmath(\thetak+\pmb{\nu}(\thetako-\thetak))})<1$}
The associated pseudo-code is provided by Algorithm \ref{alg2}, which iterates incumbent points bypassing any line search. In contrast to Algorithm \ref{alg1}, the convergence of Algorithm \ref{alg2} to at least a locally optimal solution of (\ref{n2}) can be readily proved \cite{BM93}.

{\bf Remark.} To efficiently find a reasonable $\alpha$ in (\ref{n1}), we  rely on the following procedure.
Initialize the procedure by using a sufficiently large $\alpha^{(0)}$, solve (\ref{n4}) and update
$\alpha^{(\kappa+1)} = \alpha^{(\kappa)}/10$ until not arrive at $g(e^{\jmath\thetako})\leq g(e^{\jmath\thetak})$.

We also propose an alternative full step descent procedure for (\ref{p4}) by addressing the following problem of perturbed optimization:
\begin{equation}\label{pp4}
\min_{\btheta} f_{\alpha}(e^{\jmath\btheta})\triangleq \la  \left([\clH(e^{\jmath\btheta})]^2+\alpha I_K\right)^{-1}\ra
\end{equation}
for a sufficient small $\alpha>0$. Using the matrix inverse formula (\ref{iml}) of (see e.g. \cite{ZD99}):
\begin{eqnarray}
 &&\!\left([\clH(e^{\jmath\btheta})]^2\!+\!\alpha I_K\right)^{-1} \nonumber\\&=& -
 \alpha^{-2}\clH(e^{\jmath\btheta}) \left(I_M\!+\!\alpha^{\!-1\!}[\clH^H(e^{\jmath\btheta})]^2\right)^{-1}\clH^H(e^{\jmath\btheta})\nonumber\\ && +\alpha^{-1}I_K \nonumber\\
 &=& - \alpha^{-1}\clH(e^{\jmath\btheta}) \left(\alpha I_M\!+\!\clH^H(e^{\jmath\btheta})\clH(e^{\jmath\btheta})\right)^{-1}\clH^H(e^{\jmath\btheta}) \nonumber \\
 &&+ \alpha^{-1}I_K,\label{iml}
\end{eqnarray}
the problem in (\ref{pp4}) may be shown to be equivalent to
\begin{equation}\label{pp5}
\max_{\btheta}\ g_{\alpha}(e^{\jmath\btheta})\triangleq \la\clH(e^{\jmath\btheta}) \left(\alpha I_M+[\clH^H(e^{\jmath\btheta})]^2\right)^{-1}\clH^H(e^{\jmath\btheta})\ra.
\end{equation}
Again, let $\thetak$ be a point found during the $(\kappa-1)$-st iteration.
Exploiting the inequality (\ref{fund1}) in the Appendix yields (\ref{add1})
\begin{figure*}[t]
\begin{eqnarray}
 g_{\alpha}(e^{\jmath\btheta})&\geq&-\alpha  \la[\clH^H(e^{\jmath\thetak})]^2 [\Psik]^2 \ra+ 2\Re\{\la\clH(e^{\jmath\btheta}) \Psik \clH^H(e^{\jmath\thetak})\ra \} -
 \la [\clH^H(e^{\jmath\btheta})]^2[\Psik \clH^H(e^{\jmath\thetak})]^2  \ra\nonumber\\
 &=&\ds\ak+
 2\Re\{ \sum_{n=1}^Ne^{\jmath\theta_n}\bk(n)\}+(e^{\jmath \btheta})^H\clCk e^{\jmath \btheta}\nonumber\\
 &\geq&\ds\ak+
 2\Re\left\{ \sum_{n=1}^Ne^{\jmath\theta_n}\left(\bk(n)-
 \sum_{m=1}^Ne^{-\jmath\thetak_m}\clCk(m,n)+\lambda_{\max}(\clCk)e^{-\jmath\thetak_n}\right)\right\}
\nonumber
-(e^{\jmath\thetak})^H\clCk e^{\jmath\thetak}\\&&-2\lambda_{\max}(\clCk)N\nonumber\\
 &\triangleq&\gk_{\alpha}(e^{\jmath\btheta}), \label{add1}
\end{eqnarray}
\hrulefill
\end{figure*}
for $\Psik\!=\!\left(\alpha I_M\!+\![\clH^H(e^{\jmath\thetak})]^2\!\right)\!^{-1}$,
$
\ak\triangleq \!-\!\alpha  \la[\clH^H\!(e^{\jmath\thetak}\!)]^2 [\Psik]^2 \ra$,
$\bk(n)\!\triangleq \!\la\clH_n\Psik \clH^H\!(e^{\jmath\thetak}\!)\ra$, $n\in\clN$,
and $\clCk(n,m)\triangleq \la \clH_n^H \clH_m [\Psik \clH^H\!(e^{\jmath\thetak}\!)]^2\ra$, $(n,m)\in\clN\times\clN$, while $\lambda_{\max}(\clCk)$ is the maximum eigenvalue of $\clCk$, which is positive because the matrix $\clCk$ is positive definite.

We thus solve the following problem to generate $\theta^{(\kappa+1)}$
\begin{equation}\label{pf1}
\max_{\btheta} \gk_{\alpha}(e^{\jmath\btheta}),
\end{equation}
 which admits the following closed-form solution similar to (\ref{p10}):
\begin{eqnarray}
\theta^{(\kappa+1)}_n&=&-\angle( \bk(n)-
 \sum_{m=1}^Ne^{-\jmath\thetak_m}\clCk(m,n) \nonumber \\
 &&+\lambda_{\max}(\clCk)e^{-\jmath\thetak_n}), n\in\clN.\label{pf2}
\end{eqnarray}
Similarly to (\ref{n5a}), we can readily show that  $g_{\alpha}(e^{\jmath\thetako})>g_{\alpha}(e^{\jmath\thetak})$ as far as $\thetako\neq \thetak$, so (\ref{pf2}) provides full step size update. A compact presentation of (\ref{pf1}) is also included in Algorithm \ref{alg2}.

Before concluding this section, observe that after designing $\theta^{opt}$, the throughput of all IUs is defined with the aid of (\ref{p1}) and (\ref{p2}) as
\begin{equation}\label{zfrate}
\ln \left(1+\frac{P}{\sigma\la \left([\clH(e^{\jmath\theta^{opt}})]^2\right)^{-1}\ra}\right).
\end{equation}

\begin{algorithm}
	\caption{ZFB full step descent algorithm} \label{alg2}
	\begin{algorithmic}[1]
	\State \textbf{Initialization:} Initial $\theta^{(0)}$.
		\State \textbf{Repeat until convergence of $\thetak$:} Generate $\thetako$ by (\ref{n5}) or
(\ref{pf2}). Set $\kappa:=\kappa+1$.
 \State \textbf{Output} $\theta^{opt}=\thetak$.
	\end{algorithmic}
\end{algorithm}
\section{RIS-aided regularized zero-forcing beamforming optimization}
Now, whenever we have $K>M$, the matrix $\clH(e^{\jmath\btheta})\clH^H(e^{\jmath\btheta})$ becomes singular and it cannot be inverted. Hence we  cannot use the ZFB of (\ref{zf1}). Instead, we consider  RZFB, under which the BS transmits
\begin{eqnarray}
\!x_I\!&\!=\!&\clH^H(e^{\jmath\btheta})\!\left([\clH(e^{\jmath\btheta})]^2\!+\!\alpha I_K\!\right)\!^{-1}\mbox{diag}[\bp_k]_{k=1,\dots, K}s\label{rzf1} \nonumber\\
&\!=\!&\!\left([\clH^H\!(e^{\jmath\btheta}\!)]^2\!+\!\alpha I_M\!\right)\!^{-1}\clH^H\!(e^{\jmath\btheta}\!)\mbox{diag}[\bp_k]_{k=1,\dots, K}s.\label{rzf2}
\end{eqnarray}
The Equation (\ref{lb5}) may be rewritten as
\begin{eqnarray}
y&=&\clH(e^{\jmath\btheta})\left([\clH^H(e^{\jmath\btheta})]^2+\alpha I_M\right)^{-1}\nonumber \\&&\clH^H(e^{\jmath\btheta})\mbox{diag}[\bp_k]_{k=1}^Ks +\bar{n}.\quad\label{lb7}
\end{eqnarray}

Thanks to the regularization of the ill-posed part only, we can design $\btheta$ separately, because the capability of RZFB actually depends on  the matrix $\clH(e^{\jmath\btheta})\left([\clH^H(e^{\jmath\btheta})]^2+\alpha I_M\right)^{-1}\clH^H(e^{\jmath\btheta})$ in
(\ref{lb7}). Note that we have:
\[
\begin{array}{lll}
&&\begin{bmatrix}I_{K}&\clH(e^{\jmath\btheta})\cr
\clH^H(e^{\jmath\btheta})&[\clH^H(e^{\jmath\btheta})]^2+\alpha I_M
\end{bmatrix} \\
&\succeq& \begin{bmatrix}I_{K}&\clH(e^{\jmath\btheta})\cr
\clH^H(e^{\jmath\btheta})&[\clH^H(e^{\jmath\btheta})]^2
\end{bmatrix}\\
&=&\begin{bmatrix}I_{K}\cr
\clH^H(e^{\jmath\btheta})\end{bmatrix}\begin{bmatrix}I_{K}&\clH(e^{\jmath\btheta})\end{bmatrix}
\succeq 0.
\end{array}
\]
Upon using the Shur complement (see e.g. \cite{ZD99}), we arrive at:
\begin{equation}\label{rgz1}
I_{K}\succeq \clH(e^{\jmath\btheta})\left([\clH^H(e^{\jmath\btheta})]^2+\alpha I_M\right)^{-1}\clH^H(e^{\jmath\btheta}).
\end{equation}
It is plausible that the more similar the matrix in the right hand side (RHS)  to the identity matrix in the left hand side (LHS), the better RZFB performs. Define an ellipsoid in $\mathbb{C}^K$:
\begin{eqnarray}\label{elip}
\clE(\btheta)&\triangleq &\{x\in \mathbb{C}^K: x^H \clH(e^{\jmath\btheta})\left([\clH^H(e^{\jmath\btheta})]^2+\alpha I_M\right)^{-1} \nonumber \\ &&\clH^H(e^{\jmath\btheta})x\leq 1\}.
\end{eqnarray}
The matrix inequality (\ref{rgz1}) shows that $\clE(\btheta)$ always contains the unit sphere:
\begin{equation}\label{elip1}
\clE(\btheta)\supset \clU\triangleq \{x\in\mathbb{C}^K:\ ||x||^2\leq 1\}.
\end{equation}
The rest of this section is devoted to the optimization of $\btheta$ based on optimizing the shape of $\clE(\btheta)$.
\subsection{Trace-maximization based algorithm}
We aim to minimize the surface of $\clE(\btheta)$, i.e.
we aim to maximize the trace of the right hand size (RHS) of (\ref{rgz1}) \cite{S93}.
As such, the problem is the same as (\ref{pp5}),
thus it may be solved  by Alg. \ref{alg2}. We repeat it here as Algorithm \ref{alg3} to emphasize that
it is specifically tailored  for RZFB.

\begin{algorithm}
	\caption{RZF full step descent algorithm for trace maximization (\ref{pp5})} \label{alg3}
	\begin{algorithmic}[1]
	\State \textbf{Initialization:} Initial $\theta^{(0)}$.
		\State \textbf{Repeat until convergence of $\thetak$:} Generate $\thetako$ by (\ref{pf2}).
 Set $\kappa:=\kappa+1$.
 \State \textbf{Output} $\theta^{opt}$.
	\end{algorithmic}
\end{algorithm}
\subsection{Determinant-maximization algorithms}
We aim to minimize the volume of the set $\clE(\btheta)\setminus \clU$, which is equivalent to  the problem \cite{S93}:
\begin{eqnarray}
&&\min_{\btheta}|I_{K}-\clH(e^{\jmath\btheta})\left([\clH^H(e^{\jmath\btheta})]^2+\alpha I_M\right)^{-1}\clH^H(e^{\jmath\btheta})|
\nonumber\\
&&\Leftrightarrow \max_{\btheta}\ \phi(e^{\jmath\btheta})\triangleq \ln| I_{K}+\frac{1}{\alpha}[\clH(e^{\jmath\btheta})]^2|,\label{pop1}
\end{eqnarray}
because according to the matrix inversion formula, we have
\begin{eqnarray}
&&I_{K}-\clH(e^{\jmath\btheta})\left([\clH^H(e^{\jmath\btheta})]^2+\alpha I_M\right)^{-1}\clH^H(e^{\jmath\btheta})\nonumber \\&=&
\left(I_{K}+\frac{1}{\alpha}[\clH(e^{\jmath\btheta})]^2\right)^{-1}.
\end{eqnarray}
The problem (\ref{pop1}) is equivalent to
\begin{equation}\label{pob1a}
\max_{\btheta} {\color{black}\varphi(e^{\jmath\btheta})}\triangleq \ln| \alpha I_{K}+[\clH(e^{\jmath\btheta})]^2|.
\end{equation}
As always, let $\thetak$ be a point found during the $(\kappa-1)$-st iteration.
\subsubsection{Step descent algorithm}
The linearization of the function $\varphi$ at $\thetak$ is formulated as:
\begin{equation}\label{lrzf1}
\varphi(e^{\jmath\thetak})-\la\clAk[\clH(e^{\jmath\thetak})]^2\ra
+\la \clAk[\clH(e^{\jmath\btheta})]^2\ra,
\end{equation}
for
\begin{equation}\label{lrzf2}
\clAk\triangleq \left(  \alpha I_{K}+[\clH(e^{\jmath\thetak})]^2  \right)^{-1}.
\end{equation}
Thus, for $\clAk$ defined by (\ref{lrzf2}) we address the problem (\ref{p6}) and then (\ref{p9}) to generate
the descent direction $\thetako$ given by  (\ref{p10}) as per Algorithm \ref{alg4}, which has to update the incumbent point with the convergence not granted.

\begin{algorithm}
	\caption{RZFB step descent algorithm for maximizing the log determinant (\ref{pob1a})} \label{alg4}
	\begin{algorithmic}[1]
	\State \textbf{Initialization:} Initial $\theta^{(0)}$ and set $\theta^{opt}=\theta^{(0)}$ and $\eta^{opt}=
\varphi(\theta^{opt})$ as the incumbent RIS and value.
		\State \textbf{Repeat until convergence of $\thetak$:} For $\clAk$ defined by (\ref{lrzf2}),
 generate $\thetako$ by (\ref{p10}).
If {\color{black}$\varphi(\thetako)<\eta^{opt}$, set $\theta^{opt}=\thetako$ and $\eta^{opt}=\varphi(\thetako)$.}
 Set $\kappa:=\kappa+1$.
 \State \textbf{Output} $\theta^{opt}$ and $\eta^{opt}$.
	\end{algorithmic}
\end{algorithm}

\subsubsection{Full step descent algorithm}
We also can use the inequality (\ref{fund5}) to address (\ref{pop1}) as (\ref{add3}),
\begin{figure*}[t]
\begin{eqnarray}
\phi(e^{\jmath\btheta})&\geq&\phi(e^{\jmath\thetak})-||\clH(e^{\jmath\thetak})||^2+\frac{1}{\alpha}\left[2\Re\{\clH(e^{\jmath\btheta})\clH^H(e^{\jmath\thetak})\}\right.\nonumber\\
&&\left.-\la \clH(e^{\jmath\thetak})(\alpha I_m+[\clH^H(e^{\jmath\thetak})]^2)^{-1}\clH^H(e^{\jmath\thetak}),[\clH(e^{\jmath\btheta})]^2\ra   \right]\nonumber\\
&=&\ak+\frac{1}{\alpha}\left[2\Re\{\clH(e^{\jmath\btheta})\clH^H(e^{\jmath\thetak})\}-
\la [\Psik\clH(e^{\jmath\btheta})]^2\ra   \right]\nonumber\\
&=&\ds\ak+\frac{1}{\alpha}\left[
 2\Re\{ \sum_{n=1}^Ne^{\jmath\theta_n}\bk(n)\}-(e^{\jmath \btheta})^H\clCk e^{\jmath \btheta}\right]\nonumber\\
 &\geq&\ds\ak+\frac{1}{\alpha}\left[
 2\Re\left\{ \sum_{n=1}^Ne^{\jmath\theta_n}\left(\bk(n)-
 \sum_{m=1}^Ne^{-\jmath\thetak_m}\clCk(m,n)+\lambda_{\max}(\clCk)e^{-\jmath\thetak_n}\right)\right\}\right.\nonumber\\
 &&\left.-(e^{\jmath\thetak})^H\clCk e^{\jmath\thetak}-2\lambda_{\max}(\clCk)N\right]\nonumber\\
 &\triangleq&\phik(e^{\jmath\btheta}), \label{add3}
\end{eqnarray}
\hrulefill
\end{figure*}
for $\ak\triangleq \phi(e^{\jmath\thetak})-||\clH(e^{\jmath\thetak})||^2$,
$\Psik\triangleq (\alpha I_m+[\clH^H(e^{\jmath\thetak})]^2)^{-1/2}\clH^H(e^{\jmath\thetak})$,
$\bk(n)\triangleq \la\clH_n\clH^H(e^{\jmath\thetak})\ra, n\in\clN$,
$\clCk(n,m)\triangleq \la {\color{black}\clH_m}\clH_n^H [(\Psik)^H]^2\ra$,
$(n,m)\in\clN\times\clN$,
and $\lambda_{\max}(\clCk)$ is the maximum eigenvalue of $\clCk$, which is positive because
the matrix $\clCk$ is positive definite.

We thus solve the following problem to generate a better point $\theta^{(\kappa+1)}$
\begin{equation}\label{rpf1}
\max_{\btheta} \phik(e^{\jmath\btheta}),
\end{equation}
which admits the following closed-form solution similar to (\ref{p10}):
\begin{eqnarray}\label{rpf2}
\theta^{(\kappa+1)}_n&=&-\angle( \bk(n)-
 \sum_{m=1}^Ne^{-\jmath\thetak_m}\clCk(m,n) \nonumber \\ &&+\lambda_{\max}(\clCk)e^{-\jmath\thetak_n}), n\in\clN.
\end{eqnarray}

\begin{algorithm}
	\caption{RZF full step descent algorithm for maximizing the log determinant (\ref{pob1a})} \label{alg5}
	\begin{algorithmic}[1]
	\State \textbf{Initialization:} Initial $\theta^{(0)}$.
		\State \textbf{Repeat until convergence of $\thetak$:} Generate a better point $\thetako$ by (\ref{rpf2}).
Set $\kappa:=\kappa+1$.
 \State \textbf{Output} $\theta^{opt}=\thetak$.
	\end{algorithmic}
\end{algorithm}

Before concluding this section, let us mention that after designing $\theta^{opt}$, we insert it into (\ref{lb7}) to
consider the problem  of power allocation $\bp_k$, $k=1,\dots, K$ for max-min users rate optimization. However, we
will treat it as a particular case of the problems in the next section.
\section{ZFB Applications to RIS-aided integrated data and energy delivery}
Now, in addition to IUs we consider a scenario with the BS serving also $K_E$ EUs
$e_{\ell}$, $\ell\in\clK_E\triangleq \{1,\dots, K_E\}$, which are located near the BS
to harvest energy from the BS. We employ the transmit-TS technique,
 under which the first time-slot fraction $1/t_1$ is used for energy delivery and the second time-slot fraction $1/t_2$ is used for information delivery.

\begin{figure}[!htb]
\centering
\includegraphics[width=8cm]{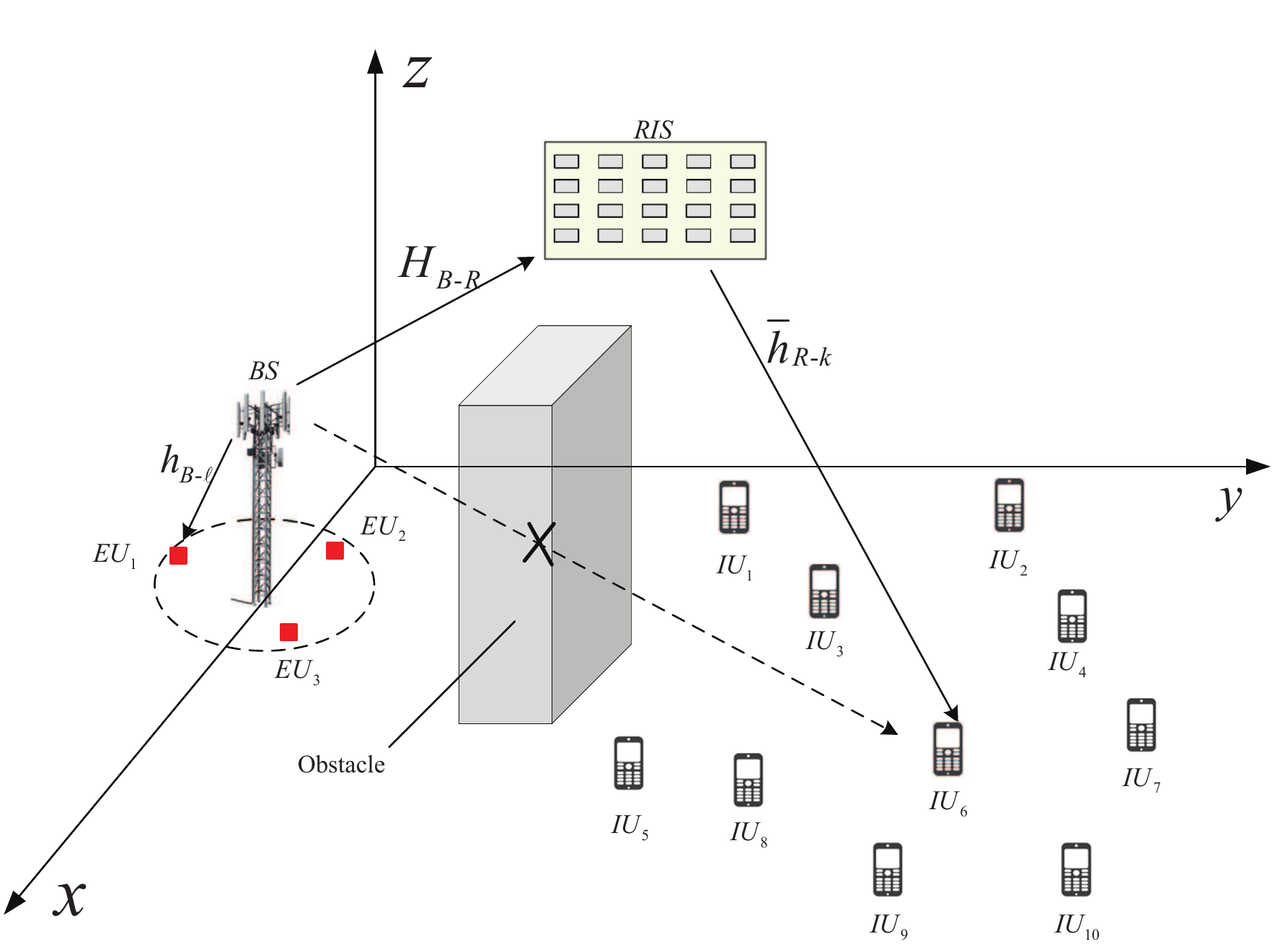}
\caption{Scenario setup for integrated data and energy networking.}\label{s1}
\end{figure}

\subsection{Energy delivery during $1/t_1$}
{\color{black} Let us assume that the LoS channel spanning from the BS to EU $e_\ell$ is $\tilde{h}_{B-e_\ell}\triangleq \sqrt{\beta^E_{B-e_\ell}}h_{B-e_\ell}\in\mathbb{C}^{1\times M}$, where $\sqrt{\beta^E_{B-e_\ell}}$ models both the path-loss and the large-scale fading of the LoS component, where the entries of $h_{B-e_\ell}$ are $\clC(0,1)$,  to modelling the small-scale fading. }

The signal received at EU $e_\ell$  is
\begin{equation}\label{eu1}
y_{e_\ell}=\tilde{h}_{B-e_\ell}x_E,
\end{equation}
where $x_E\in\mathbb{C}^M$ is the transmitted signal carrying the energy. Note that in (\ref{eu1}) we ignore
the background noise, as its power is negligible for EH. Inspired by  \cite{Yuetal20tvt}, conjugate energy beamforming is used, so we have
\[
x_E=\sum_{\ell=1}^{K_E}\tilde{h}^H_{B-e_\ell}\sqrt{\bx_\ell}\delta_{\ell},
\]
where $\delta_\ell\in{\cal C}(0,1)$, which is the energy symbol.  The power of the transmit energy signal is
\[
\pi_E(\bx)=\sum_{\ell=1}^{K_E}||\tilde{h}_{B-e_\ell}||^2\bx_\ell.
\]
The energy harvested by EU $e_\ell$ during $1/t_1$ is $\zeta \pi_{\ell}(\bx)$ with
\begin{equation}\label{eu1a}
\pi_{\ell}(\bx)\triangleq \frac{\sum_{\ell'=1}^{K_E}|\la \tilde{h}_{B-e_\ell},\tilde{h}_{B-e_{\ell'}}\ra|^2\bx_{\ell'}}{t_1},
\end{equation}
{\color{red}while} $\zeta$ is the efficiency of energy conversion, which is set to $0.5$ in this paper.
\subsection{Information transmission during $1/t_2$}
The information transmission is implemented during the time-slot fraction $1/t_2$, with the signal received at IU $k$ given by Equation (\ref{lb3}), while the corresponding multi-input multi-output (MIMO) equation is given by (\ref{lb5}). Since the specific design of PRCs has no impact on the  EH performance, we insert  $\theta^{opt}$  found in the previous sections into the equation (\ref{lb3}) and (\ref{lb5}) and also into (\ref{zf1}) and (\ref{rzf1})
for ZFB or RZF beamforming, respectively. When the ZF beamformer of(\ref{zf1}) is used in conjunction with $\bp_k\equiv \bp_0$,
the throughput at IU $k$ is expressed  according to (\ref{p1}) as
\begin{equation}\label{zps3}
r_0(\bp_0)=\ln\left(1+\bp^2_0/\sigma\right),
\end{equation}
and the power used for information transmission according to (\ref{p2}) is given by
\begin{equation}
a_{zf}\bp_0^2\label{zipo}
\end{equation}
for
\begin{equation}\label{zipo1}
a_{zf}\triangleq \la \left([\clH(e^{\jmath\theta^{opt}})]^2\right)^{-1}\ra.
\end{equation}
\subsection{Optimal energy and ZF information beamforming}
Here we consider the IUs' max-min throughput optimization problem subject to the QoES in terms of the EUs' harvested energy rate formulated as:
\begin{subequations}\label{m0}
\begin{eqnarray}
\max_{\bp_0,\bx\in\mathbb{R}_+^{K_E},\gamma,\bt=(t_1,t_2)^T\in\mathbb{R}^2_{+}}\ \gamma \quad\mbox{s.t.}\label{m0a}\\
\frac{\pi_E(\bx)}{t_1}+\frac{a_{zf}\bp_0^2}{t_2}\leq P,\label{m0b}\\
 \pi_E(\bx)\leq 3P, a_{zf}\bp_0^2\leq 3P,\label{m0c}\\
\frac{1}{t_1}+\frac{1}{t_2}\leq 1,\label{m0d}\\
\sum_{\ell'=1}^{K_E}|\la \tilde{h}_{B-e_\ell}\tilde{h}^H_{B-e_{\ell'}}\ra|^2\bx_{\ell'}\geq t_1e_{\min}/\zeta, \ell\in\clK_E,\label{m0e}\\
r_0(\bp_0)\geq \gamma t_2, \label{m0f}
\end{eqnarray}
\end{subequations}
where $e_{\min}$ is the harvested energy threshold.
The slack variable $\gamma$ is introduced in (\ref{m0a}) and (\ref{m0f}) to reflect the IUs' minimal throughput; (\ref{m0b}) is the total transmit power constraint under a given budget $P$ and (\ref{m0c}) is a physical
transmission constraint; (\ref{m0d}) restricts the energy and information transfer to a specific time slot, and (\ref{m0e}) represents the energy constraint of EUs in terms of their minimal required energy, which in fact reflects the following constraint:
\begin{equation}\label{m0ee}
\pi_{\ell}(\bx)\geq e_{\min}/\zeta, \ell\in\clK_E,
\end{equation}
with $\pi_{\ell}(\bx)$ defined in (\ref{eu1a}).

In the problem (\ref{m0}), the constraints (\ref{m0c})-(\ref{m0e}) are convex but
the constraints (\ref{m0b}) and (\ref{m0f}) are not, making (\ref{m0}) a nonconvex problem.
We now develop inner convex approximations for these nonconvex  constraints
to propose a path-following algorithm for computing (\ref{m0}).

Let $(\pk_0,\xk,\tk,\gammak)$ be a feasible point for (\ref{m0}) that is found from the $(\kappa-1)$-st iteration.
Then upon using the following inequality
\begin{equation}\label{pie}
\pi_E(\bx)\leq \paik_E(\bx)\triangleq \frac{1}{2}\left(\frac{\pi^2_E(\bx)}{\pi_E(\xk)} + \pi_E(\xk) \right),
\end{equation}
the nonconvex  constraint (\ref{m0b}) is innerly approximated by
\begin{equation}
\frac{\paik_E(\bx)}{t_1}+\frac{a_{zf}\bp_0^2}{t_2}\leq P.\label{m0e1}
\end{equation}
Using the inequality (\ref{fund4}) yields the following concave quadratic minorant of $r_0(\bp_0)$ in the LHS of (\ref{m0f}):\footnote{$r_0(\bp_0)\geq \rk_0(\bp_0)$}
\begin{eqnarray}
\rk_0(\bp_0)&\triangleq & \rk_0(\pk_0)-\frac{(\pk_0)^2}{\sigma}+ 2\frac{\pk_0}{\sigma}\bp_0 \nonumber \\
&&-\frac{(\pk_0)^2}{\sigma\left((\pk_0)^2+\sigma\right)}\left(\bp_0^2+\sigma\right).
\label{rk0}
\end{eqnarray}
Meanwhile, the RHS of (\ref{m0f}) is upper bounded as follows:
\begin{equation}\label{bo1}
\gamma t_2\leq \frac{\gammak\tk_2}{4}\left(\frac{\gamma}{\gammak}+\frac{t_2}{\tk_2}\right)^2.
\end{equation}
The nonconvex constraint (\ref{m0f}) is thus innerly approximated by the following convex quadratic constraint
\begin{equation}
\rk_0(\bp_0)\geq \frac{\gammak\tk_2}{4}\left(\frac{\gamma}{\gammak}+\frac{t_2}{\tk_2}\right)^2.\label{m01b}
\end{equation}
We then solve the following convex optimization problem for generating the next feasible point $(\pko_0,\xko,\tko,\gammako)$ for (\ref{m0}):
\begin{equation}\label{m01}
\max_{\bp_0,\bx,\gamma,\bt=(t_1,t_2)^T\in\mathbb{R}^2_{+}}\ \gamma \quad\mbox{s.t.}\quad
(\ref{m0c}),(\ref{m0d}), (\ref{m0e}), (\ref{m0e1}), (\ref{m01b}).
\end{equation}
As this convex problem involves $m_c= K_E+ 3$ decision variables and $n_v = 5$ quadratic constraints, its computational complexity is on the order of \cite{Peaucelle-02-A}
\begin{equation}\label{coc}
{\cal O}[m_c^{2.5}(n_v^2+m_c)].
\end{equation}
As $(\pko_0,\xko,\tko,\gammako)$ is the optimal solution of (\ref{m01}), while  $(\pko_0,\xk,\tk,\gammak)$ is
its feasible point, it follows that
\begin{equation}\label{b1}
\gammako>\gammak,
\end{equation}
provided that $(\pko_0,\xko,\tko,\gammako)\neq (\pk_0,\xk,\tk,\gammak)$, i.e. $(\pko_0,\xko,\tko,\gammako)$ is a better
feasible point than  $(\pk_0,\xk,\tk,\gammak)$. The sequence $\{(\pk_0,\xk,\tk,\gammak)\}$ of improved feasible points
for (\ref{m0}) converges at least to a locally optimal solution of (\ref{m0}). As analyzed in \cite{Naetal17}, this locally optimal solution often turns out to be the globally optimal one. Algorithm \ref{zfa} provides the pseudo-code for the procedure iterating (\ref{m01}).

\begin{algorithm}
	\caption{ZF Path-following algorithm for (\ref{m0})} \label{zfa}
	\begin{algorithmic}[1]
	\State \textbf{Initialization:} Randomly generate a feasible point
$(p^{(0)}_0,x^{(0)}, t^{(0)}, \gamma^{(0)})$ for (\ref{m0}). Set $\kappa=0$.
		\State \textbf{Repeat until convergence of $\gammak$:} Generate $(\pko_0,\xko,\tko,\gammako)$
 by solving the convex problem (\ref{m01}).
Set $\kappa:=\kappa+1$.
 \State \textbf{Output} $(\pk_0,\xk,\tk,\gammak)$.
	\end{algorithmic}
\end{algorithm}
\section{RZFB applications to RIS-aided integrated data and energy delivery}
\subsection{The conventional RZFB}
When the RZFB (\ref{rzf1}) is used, Equation  (\ref{lb7}) becomes
\begin{eqnarray}
y_k&=&\bar{h}_{B-k}\bar{\clH}^{rz}\sum_{j=1}^K\bar{h}^H_{B-j}\bp_js_j+\bar{n}_k\label{lb8a}\\
&=&\sum_{j=1}^K\bar{h}_{kj}\bp_js_j+\bar{n}_k,\label{lb8b}
\end{eqnarray}
for
\begin{equation}\label{topt1}
\begin{array}{c}
\bar{\clH}^{rz}\triangleq \left(\clH^H(\theta^{opt})\clH(\theta^{opt})
+\alpha I_M\right)^{-1}\\
\bar{h}_{B-j}\triangleq h_{B-j}(\theta^{opt}), j=1,\dots, K.
\end{array}
\end{equation}
and
\begin{equation}\label{top1b}
\bar{h}_{kj}\triangleq \bar{h}_{B-k}\bar{\clH}^{rz}\bar{h}^H_{B-j}.
\end{equation}
The throughput at IU $k$ is expressed as:
\begin{equation}\label{ps3}
r_k(\bp)=\ln\left(1+|\bar{h}_{kk}|^2\bp^2_k\left(\sum_{j\neq k}^K|\bar{h}_{kj}|^2\bp^2_j+\sigma\right)^{-1}  \right).
\end{equation}
The transmit power apportioned for information delivery is
\begin{equation}
\pi_I(\bp)\triangleq \sum_{j=1}^K||\bar{\clH}^{rz}\bar{h}^H_{B-j}||^2\bp^2_j.\label{ipo}
\end{equation}
Thus we consider the following problem of the IUs' max-min throughput optimization subject to the QoES in terms of the EUs' harvested energy thresholds:
\begin{subequations}\label{mpgs}
\begin{eqnarray}
\max_{\bp,\bx,\gamma,\bt=(t_1,t_2)^T\in\mathbb{R}^2_{+}}\ \gamma \quad\mbox{s.t.}\quad (\ref{m0d}), (\ref{m0e}),\label{mpgsa}\\
\frac{\pi_E(\bx)}{t_1}+\frac{\pi_I(\bp)}{t_2}\leq P,\label{mpgsb}\\
 \pi_E(\bx)\leq 3P, \pi_I(\bp)\leq 3P,\label{mpgsc}\\
r_k(\bp)\geq \gamma t_2, k\in\clK,\label{mpgsf}
\end{eqnarray}
\end{subequations}
where like their counterparts in (\ref{m0}),  $\gamma$ in (\ref{mpgsa})
and (\ref{mpgsf}) is a slack variable to express the IUs minimal throughput, (\ref{mpgsb}) and
(\ref{mpgsc}) are respectively  the total power transmit constraint under the budget $P$ and a physical
transmission constraint, while  as before, (\ref{m0d}) restricts the energy and information transfer within  a time slot,
and  (\ref{m0e}) is the energy constraint of EUs in terms of their minimal required energy.

To propose a path-following algorithm for computing (\ref{mpgs}), we have to develop inner approximations for
its  nonconvex  constraints (\ref{mpgsb}) and (\ref{mpgsf}).

Let $(\pk,\xk,\tk,\gammak)$ be a feasible point for (\ref{mpgs}) that is found from the $(\kappa-1)$-st iteration.
Based on the inequality (\ref{pie}), the nonconvex  constraint (\ref{mpgsb}) is innerly approximated by
\begin{equation}
\frac{\paik_E(\bx)}{t_1}+\frac{\pi_I(\bp)}{t_2}\leq P.\label{mpgse1}
\end{equation}
Using the inequality (\ref{fund4}) yields the following concave quadratic minorant of $r_k(\bP)$ in the LHS of (\ref{mpgsf}):
\begin{eqnarray}
r_k^{(k)}(\bP)\triangleq  \ttak + 2\ttbk\bp_k -\ttck\sum_{j=1}^K|\bar{h}_{kj}|^2\bp^2_j,\label{rkk}
\end{eqnarray}
where
\begin{gather}
\ttak_k = r_k(\pk) - |\bar{h}_{kk}|^2(\pk_k)^2\left(\sum_{j\neq k}|\bar{h}_{kj}|^2(\pk_j)^2+\sigma\right)^{-1} \nonumber \\
-\sigma\ttck_k,\qquad\qquad\qquad\qquad\qquad\qquad\qquad  \nonumber \\
\ttbk_k = |\bar{h}_{kk}|^2\pk_k\left(\sum_{j\neq k}|\bar{h}_{kj}|^2(\pk_j)^2+\sigma\right)^{-1},\nonumber\\
\ttck_k \!= \!\left(\sum_{j\neq k}|\bar{h}_{kj}|^2\!(\pk_j\!)^2\!+\!\sigma\right)^{-1} \!- \!\left(\sum_{j=1}^K|\bar{h}_{kj}|^2\!(\pk_j\!)^2\!+\!\sigma\right)^{-1}.\nonumber
\end{gather}

By (\ref{bo1}) and (\ref{rkk}), the nonconvex constraint (\ref{mpgsf}) is
innerly approximated by the following convex quadratic constraint
\begin{equation}
r_k^{(k)}(\bp)\geq \frac{\gammak\tk_2}{4}\left(\frac{\gamma}{\gammak}+\frac{t_2}{\tk_2}\right)^2,
k\in\clK.\label{mpgs1b}
\end{equation}
We then solve the following convex optimization problem for generating the next feasible point $(\pko,\xko,\tko,\gammako)$ for (\ref{mpgs}):

\begin{equation}\label{mpgs1}
\max_{\bp,\bx,\gamma,\bt=(t_1,t_2)^T\in\mathbb{R}^2_{+}}\ \gamma \quad\mbox{s.t.}\quad
(\ref{m0d}), (\ref{m0e}), (\ref{mpgsc}), (\ref{mpgse1}), (\ref{mpgs1b}).
\end{equation}
The computational complexity order of this convex problem is given by (\ref{coc}) where we have  $n_v=K + K_E+ 3$ and $m_c= K+4$. As $(\pko,\xko,\tko,\gammako)$ is the optimal solution of (\ref{mpgs1}) while  $(\pko,\xk,\tk,\gammak)$ is
its feasible point,  (\ref{b1}) is satisfied, provided that $(\pko,\xko,\tko,\gammako)\neq (\pk,\xk,\tk,\gammak)$, i.e. $(\pko,\xko,\tko,\gammako)$ is a better
feasible point than  $(\pk,\xk,\tk,\gammak)$. The sequence $\{(\pk,\xk,\tk,\gammak)\}$ of improved feasible points
for (\ref{mpgs}) converges at least to a locally optimal solution of (\ref{mpgs}). Algorithm \ref{apgs} provides
the pseudo-code for solving (\ref{mpgs}) by iterating the convex problem (\ref{mpgs1}).

\begin{algorithm}
	\caption{Conventional RZF Path-following algorithm for (\ref{mpgs})} \label{apgs}
	\begin{algorithmic}[1]
	\State \textbf{Initialization:} Randomly generate a feasible point
$(p^{(0)},x^{(0)}, t^{(0)}, \gamma^{(0)})$ for (\ref{mpgs}). Set $\kappa=0$.
		\State \textbf{Repeat until convergence of $\gammak$:} Generate $(\pko,\xko,\tko,\gammako)$
 by solving the convex problem (\ref{mpgs1}).
 Set $\kappa:=\kappa+1$.
 \State \textbf{Output} $(\pk,\xk,\tk,\gammak)$.
	\end{algorithmic}
\end{algorithm}
\subsection{New RZFB}
Instead of (\ref{rzf1}), let us now design the transmit signal as
\begin{eqnarray}
x_I&=&\bar{\clH}^{rz}\sum_{j=1}^K\bar{h}^H_{B-j}[\bp_{1,j}s_j+\bp_{2,j}s_j^*]\label{igs1}
\end{eqnarray}
with $\bp_{1,j}\in\mathbb{C}$ and $\bp_{2,j}\in\mathbb{C}$, so instead of (\ref{lb8b}) the received signal by IU $k$ is
\begin{eqnarray}
y_k=\sum_{j=1}^K\bar{h}_{kj}[\bp_{1,j}s_j+\bp_{2,j}s_j^*]+\bar{n}_k.\label{igs2}
\end{eqnarray}
While $x_I$ defined by (\ref{rzf1}) is a proper Gaussian random variable with $\mathbb{E}((x_I)^2)=0$, that defined
by (\ref{igs1}) is an improper Gaussian random variable \cite{SS10} with $\mathbb{E}((x_I)^2)\neq 0$.
The  augmented form of (\ref{igs2}) is
\begin{equation}\label{igs3}
\bar{y}_k=\sum_{k=1}^K\bar{H}_{kj}V(\bp_j)\bar{s}_j+\bar{n}^A_k,
\end{equation}
where
\[
\bar{y}_k\triangleq \begin{bmatrix}y_k\cr
y_k^*\end{bmatrix}, \bar{H}_{kj}=\begin{bmatrix}\bar{h}_{kj}&0\cr
0&\bar{h}_{kj}^*\end{bmatrix}, \bar{s}_j\triangleq \begin{bmatrix}s_j\cr
s_j^*\end{bmatrix}, \bar{n}^A_k\triangleq \begin{bmatrix}\bar{n}_k\cr
\bar{n}^*_k\end{bmatrix},
\]
and
\[ V(\bp_j)\triangleq \begin{bmatrix}\bp_{1,j}&\bp_{2,j}\cr
\bp_{2,j}^*&\bp_{1,j}^*\end{bmatrix}, \bp_j\triangleq (\bp_{1,j},\bp_{2,j}).
\]
The throughput of IU $k$ is $\frac{1}{2}\rho_k(\bp)$ \cite{CT06} with
\[
\rho_k(\bp)\!\triangleq \!\ln\left|I_2\!+\![\bar{H}_{kk}V(\bp_k)]^2\left(\sum_{j\neq k}^K[\bar{H}_{kj}V(\bp_j)]^2\!+\!\sigma I_2   \right)^{-1}   \right|.
\]
The transmit power apportioned for information delivery is
\begin{equation}
\tpi_I(\bp)\triangleq \sum_{j=1}^K||\bar{\clH}^{rz}\bar{h}^H_{B-j}\begin{bmatrix}\bp_{1,j}&\bp_{2,j}\end{bmatrix}||^2.\label{igpo}
\end{equation}
Thus we consider the following problem of the IUs' max-min rate optimization subject to the QoES in terms of the EUs' harvested energy thresholds corresponding to (\ref{mpgs}):
\begin{subequations}\label{migs}
\begin{eqnarray}
\max_{\bp,\bx,\gamma,\bt=(t_1,t_2)^T\in\mathbb{R}^2_{+}}\ \gamma \quad\mbox{s.t.}\quad
(\ref{m0d}), (\ref{m0e}),
\label{migsa}\\
\frac{\pi_E(\bx)}{t_1}+\frac{\tpi_I(\bp)}{t_2}\leq P,\label{migsb}\\
 \pi_E(\bx)\leq 3P, \tpi_I(\bp)\leq 3P,\label{migsc}\\
\rho_k(\bp)\geq 2\gamma t_2, k\in\clK.\label{migsd}
\end{eqnarray}
\end{subequations}
where like their counterparts in (\ref{mpgs}), the slack variable $\gamma$ is introduced
in (\ref{migsa}) and (\ref{migsd})  to express the IUs' minimal throughput, (\ref{migsb}) and
(\ref{migsc}) are respectively the total power transmit constraint under the budget $P$ and a physical
transmission constraint, while  as before, (\ref{m0d}) restricts the energy and information transfer within a time slot,
and  (\ref{m0e}) is the energy constraint of EUs in terms of their minimal required energy.
To propose a path-following algorithm for computing (\ref{migs}), we have to develop inner approximations for
its  nonconvex  constraints (\ref{migsb}) and (\ref{migsd}).

Let $(\pk,\xk,\tk,\gammak)$ be a feasible point for (\ref{mpgs}) that is found from the $(\kappa-1)$-st iteration.

Using the inequality (\ref{fund4}) in the Appendix yields
\begin{eqnarray}
\rho_k(\bp)&\geq& \ak_k + 2\Re\{\la V^H(\pk_k)\bar{H}^H_{kk}(\Bk_k)^{-1} \bar{H}_{kk}V(\bp_k)\ra\}\nonumber \\
&& -\sum_{j=1}^K||(\Ck)^{1/2}\bar{H}_{kj}V(\bp_j)||^2 \nonumber\\
&\triangleq&\rho_k^{(k)}(\bP),\label{irkk}
\end{eqnarray}
where
\[
\begin{array}{c}
\ak_k \triangleq r_k(\pk) -\la [\bar{H}_{kk}V(\pk_k)]^2(\Bk_k)^{-1}\ra -\sigma\la \Ck_k\ra,\\
\Bk_k\triangleq   \sum_{j\neq k}^K[\bar{H}_{kj}V(\pk_j)]^2+\sigma I_2,\\
\Ck_k\triangleq  (\Bk_k)^{-1}-\left(\Bk_k+[\bar{H}_{kk}V(\pk_k)]^2 \right)^{-1}.
\end{array}
\]
We then solve the following convex  problem for generating the next better feasible point
$\left(\pko,\xko\right)$ for (\ref{migs}):
\begin{subequations}\label{migs1}
\begin{eqnarray}
\max_{\bp,\bx,\gamma,\bt=(t_1,t_2)^T\in\mathbb{R}^2_{+}}\ \gamma \nonumber \\
\mbox{s.t.}\label{migs1a}\quad
(\ref{m0d}), (\ref{m0e}), (\ref{mpgse1}), (\ref{migsb}), (\ref{migsc}), \label{migs1a}\\
\rho_k^{(k)}(\bp)\geq \frac{\gammak\tk_2}{2}\left(\frac{\gamma}{\gammak}+\frac{t_2}{\tk_2}\right)^2.\label{migs1b}
\end{eqnarray}
\end{subequations}
The computational complexity order of this convex problem is given by (\ref{coc}) for $n_v=2K + K_E+ 3$ and $m_c= K+4$. The pseudo-code for iterating (\ref{migs1}) for computing (\ref{migs}) is provided by Algorithm \ref{aigs}.

\begin{algorithm}
	\caption{New RZF Path-following algorithm for (\ref{migs})} \label{aigs}
	\begin{algorithmic}[1]
	\State \textbf{Initialization:} Randomly generate a feasible point
$(p^{(0)},x^{(0)}, t^{(0)}, \gamma^{(0)})$ for (\ref{migs}). Set $\kappa=0$.
		\State \textbf{Repeat until convergence of $\gammak$:} Generate $(\pko,\xko,\tko,\gammako)$
 by solving the convex problem (\ref{migs1}).
Set $\kappa:=\kappa+1$.
 \State \textbf{Output} $(\pk,\xk,\tk,\gammak)$.
	\end{algorithmic}
\end{algorithm}
\subsection{Notices on RZFB for information delivery only}
When $K_E=0$, i.e. there are no EUs, we use the whole time-slot for information transfer. Hence we have $t_2=1$, i.e. the problems
(\ref{mpgs}) and (\ref{migs}) are respectively reduced to
\begin{equation}\label{smpgs}
\max_{\bp}\min_{k=1,\dots, K}r_k(\bp)\quad\mbox{s.t.}\quad \pi_I(\bp)\leq P,
\end{equation}
and
\begin{equation}\label{smigs}
\max_{\bp}\min_{k=1,\dots, K}\rho_k(\bp)\quad\mbox{s.t.}\quad \tpi_I(\bp)\leq P.
\end{equation}
Algorithms \ref{apgs} and \ref{aigs} are directly adjusted for their computation. The pseudo-code is provided by
Algorithm \ref{adata}.

\begin{algorithm}
	\caption{Path-following algorithm for (\ref{smpgs})/(\ref{smigs})} \label{adata}
	\begin{algorithmic}[1]
	\State \textbf{Initialization:} Randomly generate a feasible point
$p^{(0)}$ for (\ref{smpgs})/(\ref{smigs}). Set $\kappa=0$.
		\State \textbf{Repeat until convergence of $\gammak$:} Generate $\pko$
 by solving the convex problem $\max_{\bp}\min_{k=1,\dots, K}\rk_k(\bp)\quad\mbox{s.t.}\quad \pi_I(\bp)\leq P$
(for computing (\ref{smpgs})) and  $\max_{\bp}\min_{k=1,\dots, K}\rhok_k(\bp)\quad\mbox{s.t.}\quad \tpi_I(\bp)\leq P$
(for computing (\ref{smigs}))
Set $\kappa:=\kappa+1$.
 \State \textbf{Output} $\pk$.
	\end{algorithmic}
\end{algorithm}
\section{Numerical examples}
In this section, we investigate the performance of our proposed algorithms by numerical examples.  The elements of the  BS-to-RIS LoS channel matrix are generated by $[H_{B-R}]_{n,m} = e^{j \pi \left( (n-1) \sin \bar{\theta}_n \sin \bar{\phi}_n  + (m-1) \sin  e^{\jmath \theta_n} \sin \phi_n  \right) }$, where $ e^{\jmath \theta_n}$ and $\phi_n$ are uniformly distributed as $ e^{\jmath \theta_n} \sim \mathcal{U}(0,\pi)$ and $\phi_n \sim \clU(0,2\pi)$, respectively, and $\bar{\theta}_n = \pi - {\theta}_n$ and $\bar{\phi}_n = \pi + {\phi}_n$
\cite{Nadetal20}. The normalized small-scale fading channel $h_{B-e^\ell}$ spanning from the BS to EU $\ell$ and $\bar{h}_{R-k}$ of the RIS to IU $k$  obeys Rician distribution with a K-factor of $3$ for  modeling the LoS channels. The large scale fading coefficients, $\beta_{B-R}$, $\beta_{R-k}$, and   $\beta^E_{B-e_\ell}$,  are modeled as \cite{Nadetal20,Yuetal20tvt}
\begin{subequations}  \label{simulationb}
\begin{eqnarray}
\beta_{B-R} &\!=\!& G_\text{BS} \!+ \!G_\text{RIS} \!-\! 35.9 \!- \!22 \log_{10}(d_{B-R}) \ \text{(in dB)},\nonumber\\
\!\beta_{R-k} &\!=\!& G_\text{RIS}  - 33.05 - 30 \log_{10}(d_{RIS-k}) \ \text{(in dB)},\nonumber\\
\!\beta^E_{B-e_\ell} &\!=\!& G_\text{BS}  - 30 - 20 \log_{10}(d_{B-e^\ell})  \ \text{(in dB)},\nonumber
\end{eqnarray}
\end{subequations}
where $G_\text{BS} = 5$ dBi and $G_\text{RIS} = 5$ dBi denote the antenna gain of the BS and the RIS gain, respectively, while $d_{B-R}$, $d_{RIS-k}$, and $d_{B-e^\ell}$ are the distances between the BS and RIS, the RIS and IU $k$, and the BS and EU $\ell$, respectively. The signal reflected by the RIS can be ignored for EUs, since $\beta_{B-R}\beta_{R-k} \ll \beta^E_{B-e_\ell}$.
 The spatial correlation matrix is given by $[\mathbf{R}_{RIS-k}]_{n,n'} = e^{j\pi(n-n') \sin \tilde{\phi}_k \sin \tilde{\theta}_k }$, where $\tilde{\phi}_k$ and $\tilde{\theta}$ are the azimuth and elevation angle for IU $k$, respectively. Unless otherwise stated,  $K =10$, $K_E=3$, $e_0 = -20$ dBm and $N = 100$ are used. The results are multiplied by $\log_2 e$ to convert units of nats/sec into units of bps/Hz. The convergence tolerance of the proposed algorithms is set to $10^{-3}$.  All simulations implemented on a Core i7-10875H 2.30GHz  processor.

We use the $3$D coordinates $(x,y,z)$ to locate all the objects concerned . The BS is at $(20,0,10)$, the RIS is at $(0,30,40)$.
\subsection{RIS-aided information delivery}
 There are $K=10$ IUs, which are randomly placed in a $60m \times 60 m$ area RHS of the obstacle and the RIS. Unless stated otherwise, the transmit power of  $P = 25$ dBm is used. The performance of
Algorithm \ref{alg1} is not sensitive to which step size from Eqaution (\ref{spl}), (\ref{gr1}) and (\ref{gr1b}) is used. In the simulated figures, Alg 2A and Alg 2B refer to the performance of the  full step descent algorithm \ref{alg2}  based on iterating (\ref{n5})  and (\ref{pf2}), respectively.   Alg 3, Alg 4 and Alg 5 respectively refer to the performance  of  Algorithm \ref{alg3} for the trace-maximization (\ref{pp5}), Algorithm \ref{alg4} and Algorithm \ref{alg5} for the log determinant maximization (\ref{pob1a}). ZFB random $\theta$ and  RZFB random $\theta$ respectively refer to the performance of ZFB
and RZFB under
random PRCs.

\begin{figure}[!htb]
\centering
\begin{minipage}[t]{0.48\textwidth}
\centering
\includegraphics[width=8cm]{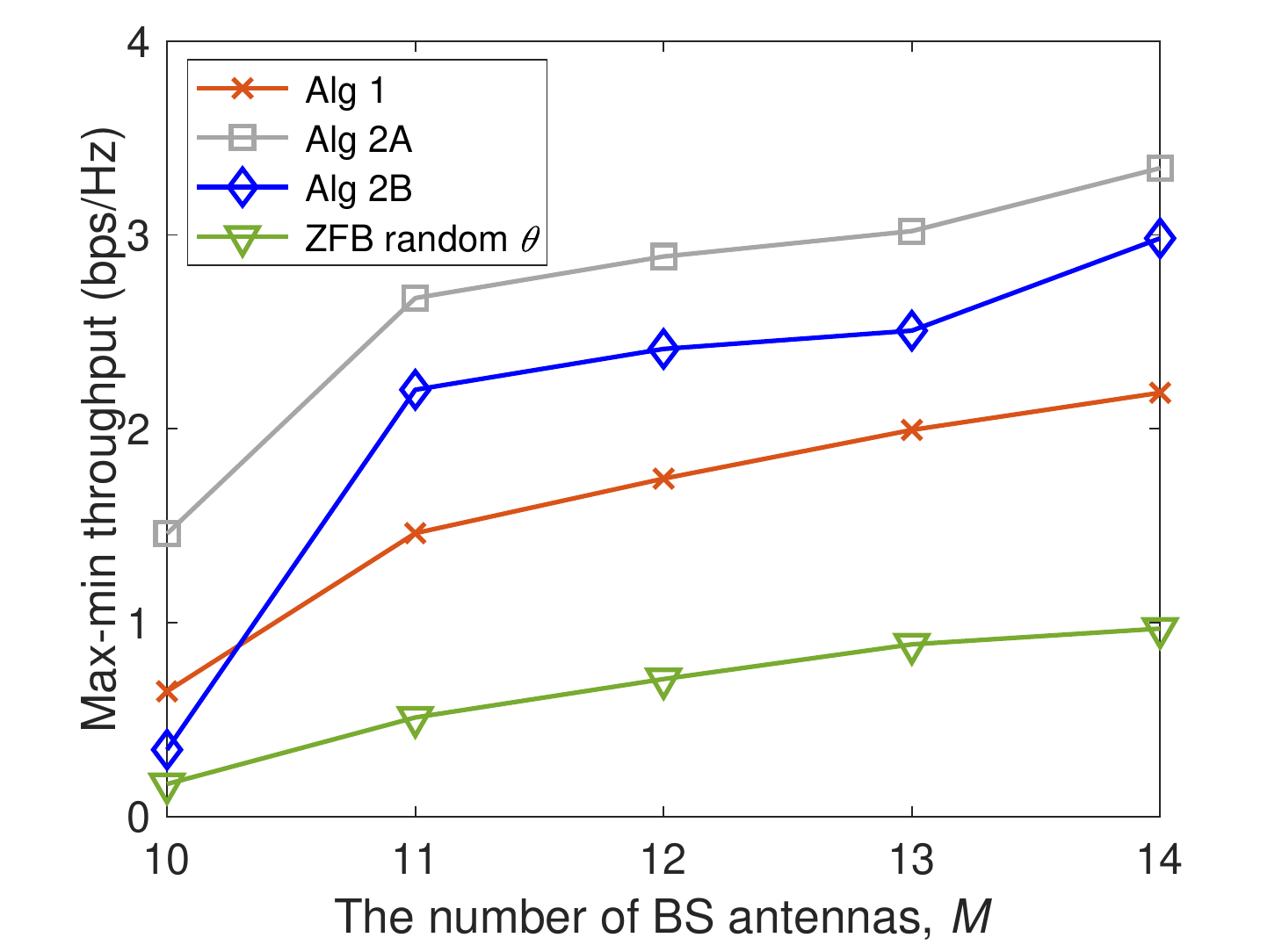}
\caption{Achievable minimum throughput vs the number of BS antennas $M$ under ZF.}\label{ZF1}
\end{minipage}
\hfill
\begin{minipage}[t]{0.48\textwidth}
\centering
\includegraphics[width=8cm]{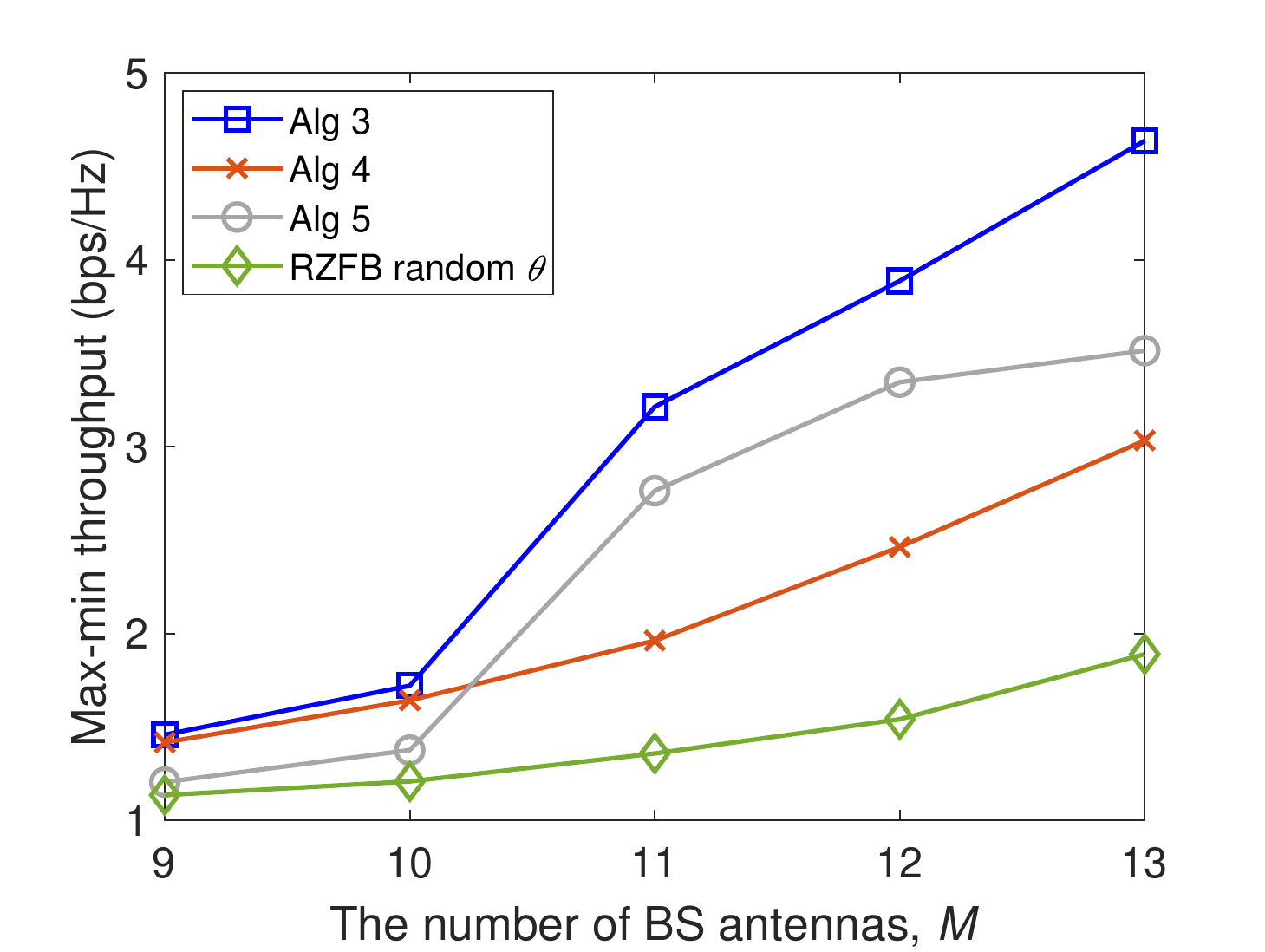}
\caption{Achievable minimum throughput vs the number of BS antenna $M$ under RZF.}\label{RZF1}
\end{minipage}
\end{figure}

Fig. \ref{ZF1} and Fig. \ref{RZF1} plot the achievable minimum throughput versus the number of BS antennas, $M$ under ZFB and RZFB, respectively.

Regarding ZFB for $M > K$, Fig. \ref{ZF1} reveals that Algorithm \ref{alg2}A outperforms
Algorithm \ref{alg2}B, and the latter  outperforms Algorithm \ref{alg1},  showing that the concave optimization reformulation (\ref{n1e}) is the best option for computing (\ref{p4}), while Algorithm \ref{alg1} of common purpose step descent is inefficient. Furthermore, the average running time for Algorithm \ref{alg1}, Algorithm \ref{alg2}A and Algorithm \ref{alg2}B are 0.24s, 0.06s and 7.96s under $N =100$, respectively.

Regarding RZFB for $M>K$, Fig. \ref{RZF1} shows that Algorithm \ref{alg3} achieves the best performance, i.e. the trace-maximization (\ref{pp5}) has a more beneficial impact on the IUs' throughput than the log determinant maximization (\ref{pob1a}). It is not surprising to see that Algorithm \ref{alg5} outperforms Algorithm \ref{alg4}
because the former iterates the incumbent points, while the latter simply provides a way to locate a beneficial direction.

The worst performance is attained by   ZFB random $\theta$ and  RZFB random $\theta$ in all figures which is a clear indication that the PRC optimization is absolutely necessary upon using RIS.

\subsection{RIS-aided information and energy delivery}
Next, we consider the problem of RIS-aided information and energy delivery by the network
of Fig. \ref{s1} with $K_E=3$ EUs randomly placed within a radius of $10$m from the BS.

Alg 1-PGS, Alg 2A-PGS, Alg 2B-PGS and ZFB random $\theta$-PGS refer to the performance of
Algorithm \ref{alg1}, Algorithm \ref{alg2}A, Algorithm \ref{alg2}B, and ZFB random $\theta$.
Alg 3-PGS, Alg 4-PGS, Alg 5-PGS, and RZFB random $\theta$-PGS refer to the performance of
Algorithm \ref{alg3}, Algorithm  \ref{alg4}, Algorithm \ref{alg5}, and RZFB random $\theta$ under
RZFB (\ref{rzf1}), while Alg 3-IGS, Alg 4-IGS and Alg 5-IGS refer to the performance of
Algorithm \ref{alg3}, Algorithm  \ref{alg4}, Algorithm \ref{alg5} under the new RZFB (\ref{igs1}).

\begin{figure}[!htb]
\centering
\begin{minipage}[t]{0.48\textwidth}
\centering
\includegraphics[width=8cm]{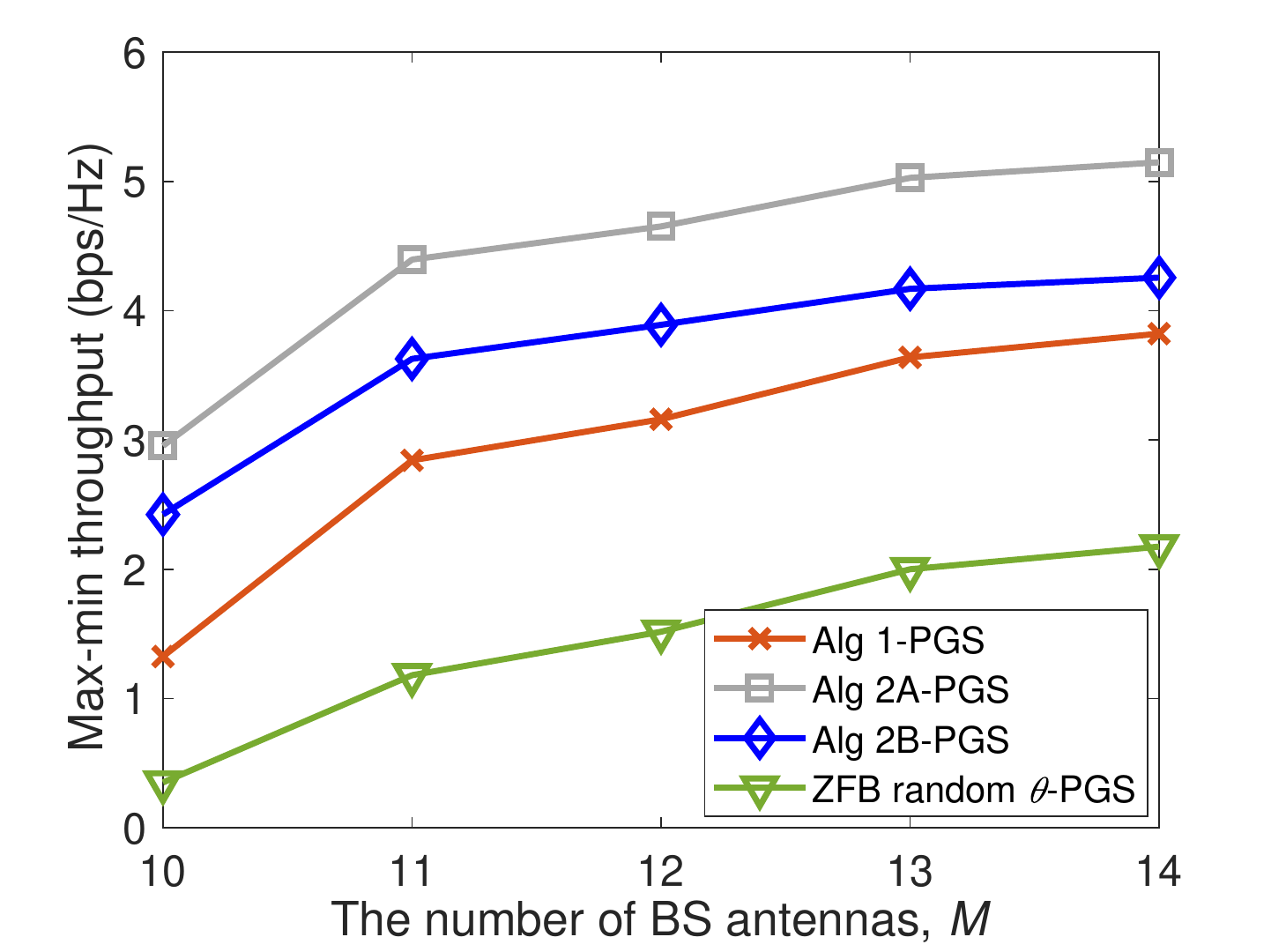}
\caption{Achievable minimum throughput vs the number of BS antennas $M$  under ZF.}\label{ZF2}
\end{minipage}
\hfill
\begin{minipage}[t]{0.48\textwidth}
\centering
\includegraphics[width=8cm]{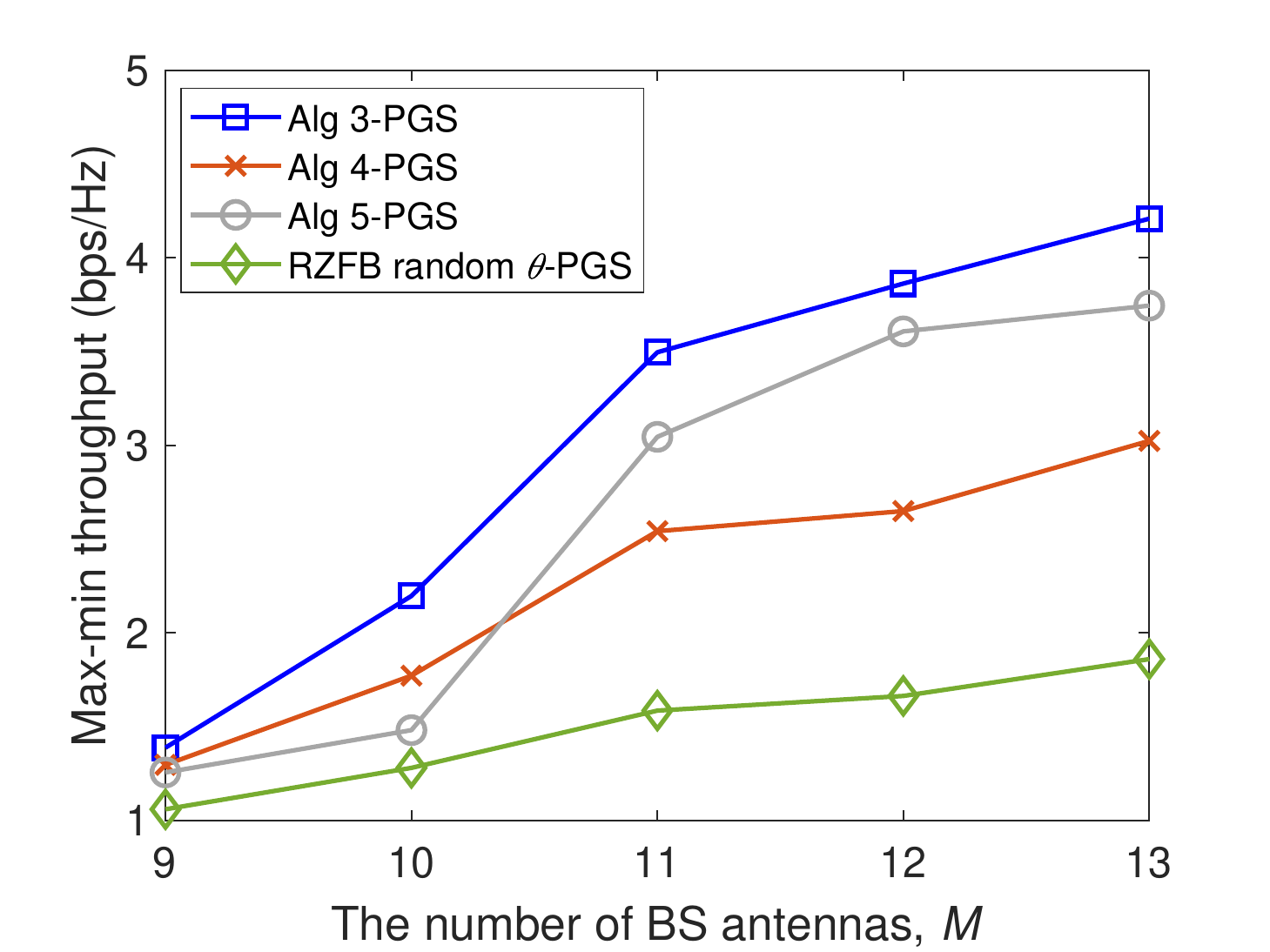}
\caption{Achievable minimum throughput vs the number of BS antennas $M$  under RZF.}\label{RZF2}
\end{minipage}
\end{figure}

\begin{figure}[!htb]
\centering
\begin{minipage}[t]{0.48\textwidth}
\centering
\includegraphics[width=8cm]{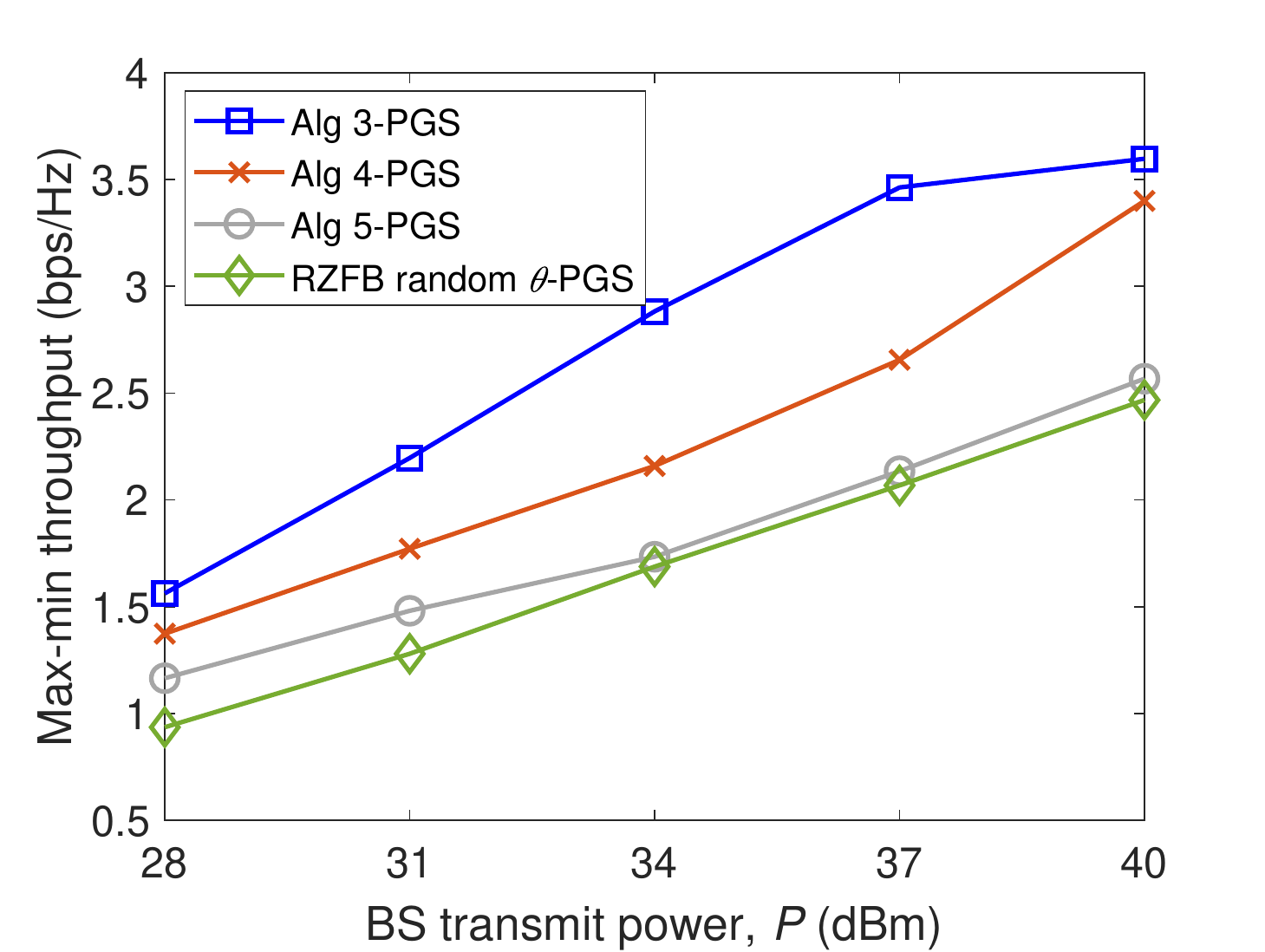}
\caption{Achievable minimum throughput for $M =10 $ under RZF vs the BS transmit power  $P$.}\label{RZF3}
\end{minipage}
\hfill
\begin{minipage}[t]{0.48\textwidth}
\centering
\includegraphics[width=8cm]{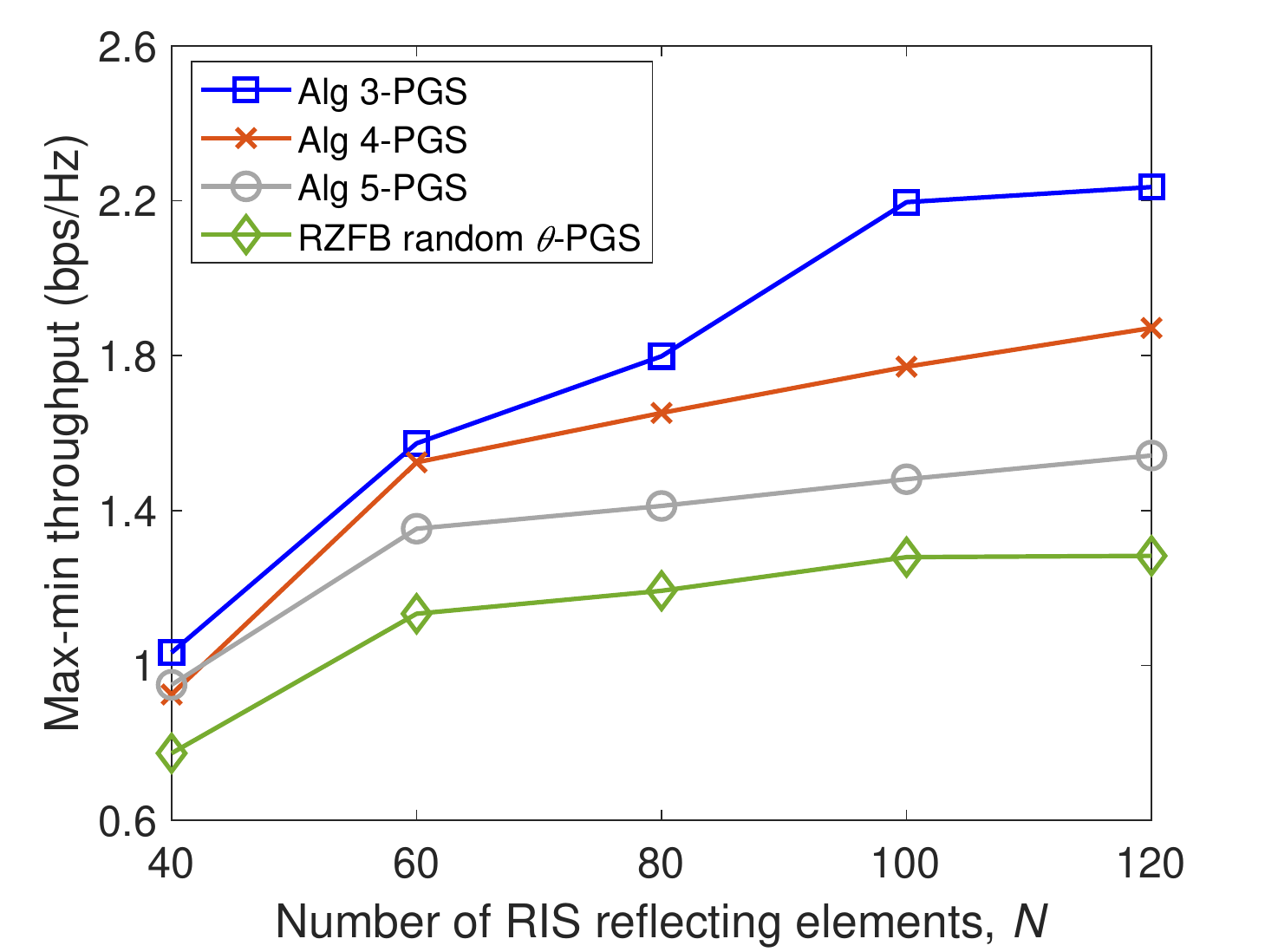}
\caption{Achievable minimum throughput for $M =10 $ with energy harvesting under RZF vs RIS for $N$ reflecting elements.}\label{RZF4}
\end{minipage}
\end{figure}

\begin{figure}[!htb]
\centering
\begin{minipage}[t]{0.48\textwidth}
\centering
\includegraphics[width=8cm]{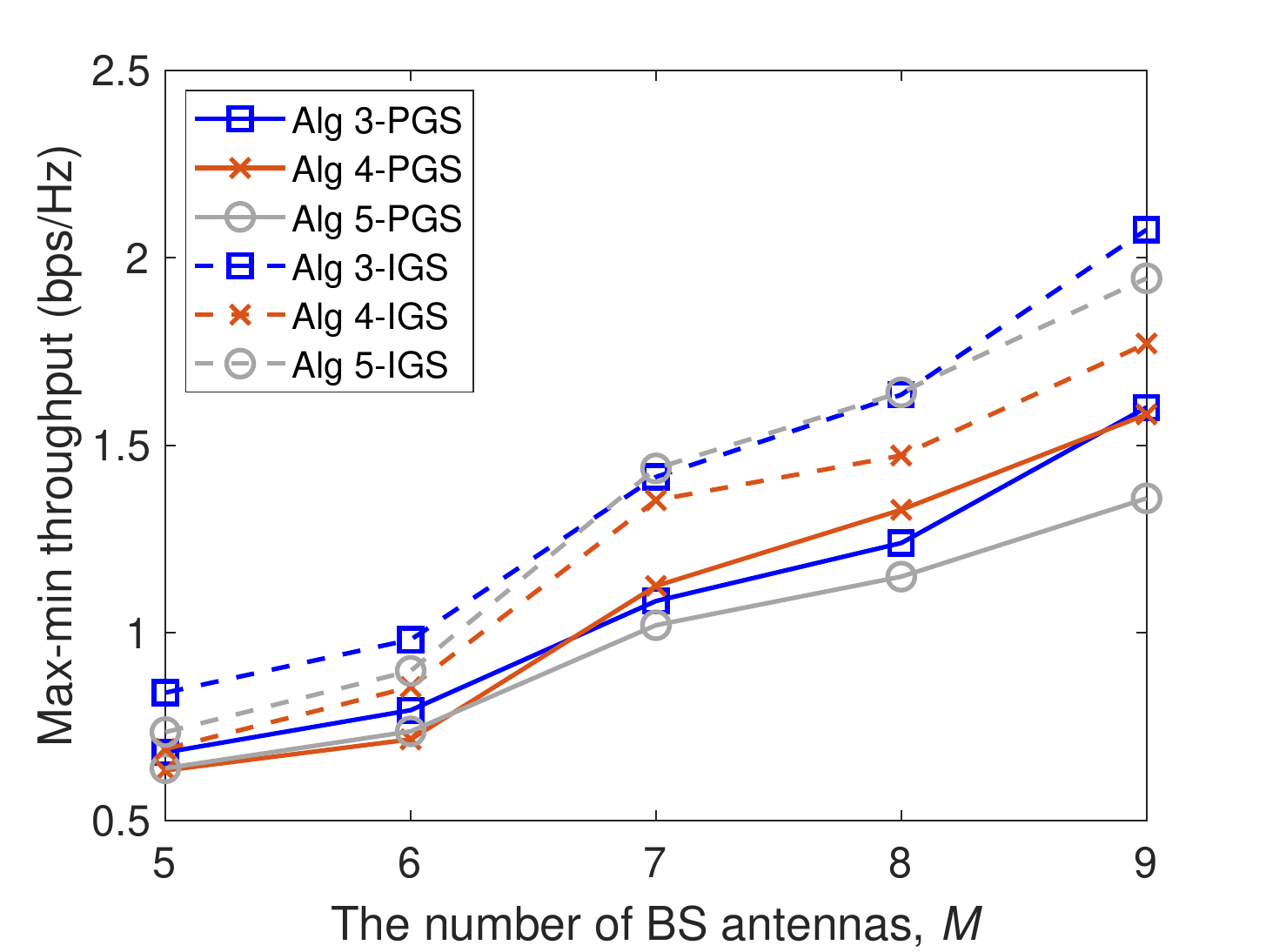}
\caption{Achievable minimum throughput under RZF vs the number of BS antennas $M$.}\label{RZF5}
\end{minipage}
\hfill
\begin{minipage}[t]{0.48\textwidth}
\centering
\includegraphics[width=8cm]{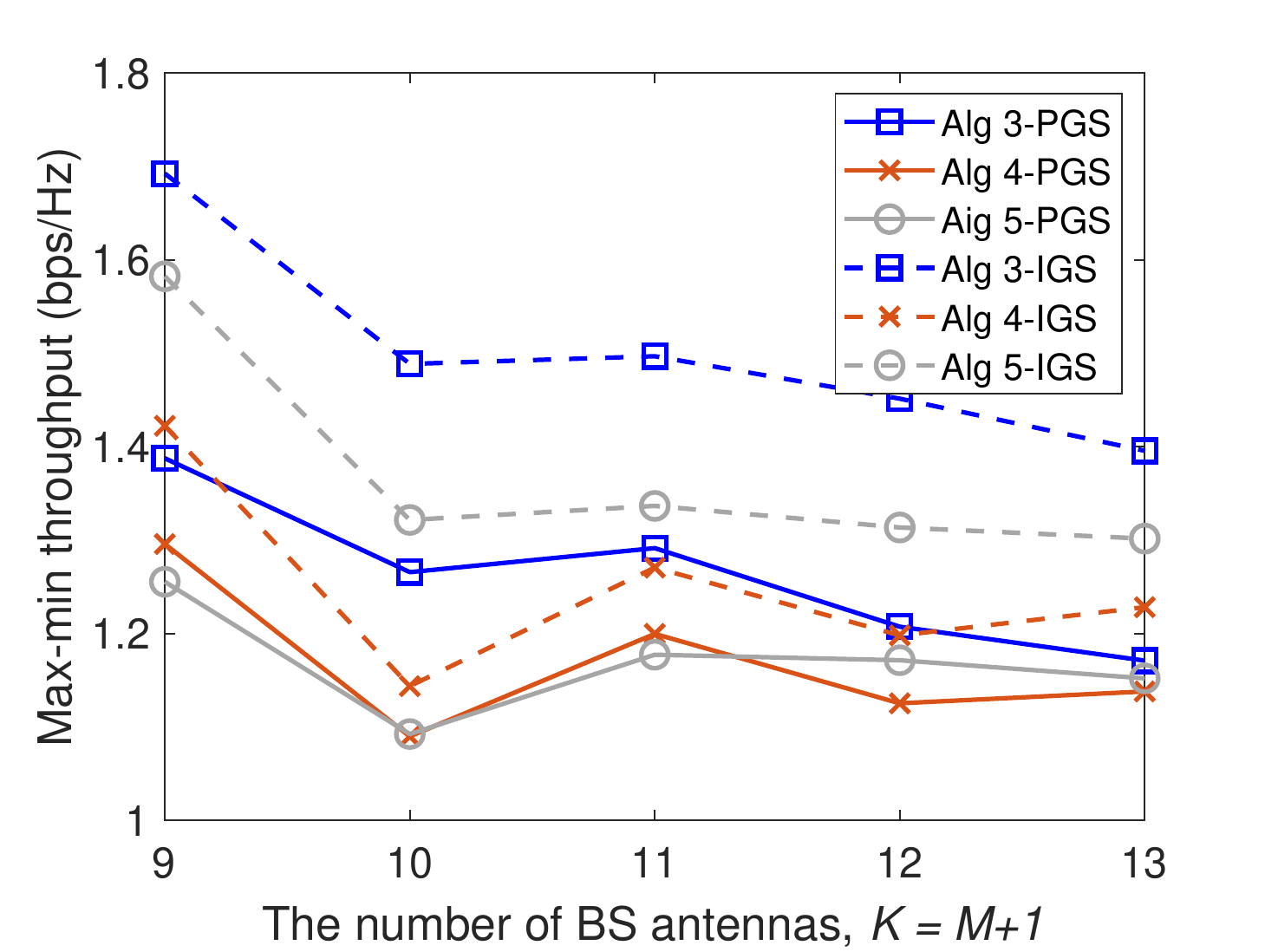}
\caption{Achievable minimum throughput under RZF for $K = M +1$ BS antennas.}\label{RZF6}
\end{minipage}
\end{figure}

The transmit power of $P = 31$ dBm is set in Fig. \ref{RZF2}-Fig. \ref{RZF4} and Fig. \ref{RZF6}, but
$P = 35$ dBm is set in Fig. \ref{RZF5} due to the relative small  numbers of BS antennas.

Fig. \ref{ZF2} and Fig. \ref{RZF2} plot the minimum achievable IU throughput versus the number
$M$ of BS antennas under ZFB and RZFB, respectively. In Fig. \ref{ZF2}, Alg 2A-PGS outperforms Alg 1-PGS and Alg 2B-PGS, Alg 2B-PGS outperforms Alg 1-PGS. Furthermore, all the proposed algorithms outperform ZFB random $\theta$-PGS. As expected, Figures  \ref{RZF1} and \ref{RZF2} exhibit similar trend.
Fig. \ref{ZF2} and Fig. \ref{RZF2} also confirm the gain achieved by optimizing the PRCs.
Furthermore, all algorithms  in Fig. \ref{ZF1}, Fig. \ref{RZF1}, Fig. \ref{ZF2} and Fig. \ref{RZF2} benefit from the spatial diversity, which is commensurate with the number BS antennas.

In Fig. \ref{RZF3}, we now examine the minimum achievable IU throughput upon varying the BS transmit power budget $P$ under RZFB for $M = 10$ . As excepted, the IUs' minimum throughput increases upon increasing the available power budget due to the availability of more power for information delivery. Naturally, beyond a certain threshold, namely $P = 40$ dBm in Fig. \ref{RZF3}, Alg 3's performance becomes saturated because the network is  interference-limited. Fig. \ref{RZF3} also shows the gap between Algorithm \ref{alg5} and RZFB under random $\theta$, which is quite narrow for $P \geq 34$ dBm because the beneficial impact of the RIS is reduced, when the power budget is increased.

Fig. \ref{RZF4} plots the achievable minimum IU throughput for $M =10 $ under RZF versus the number
$N$ of RIS reflecting elements, showing that the performance is improved upon increasing $N$.

Fig. \ref{RZF5} and Fig. \ref{RZF6} allow us to compare the performance achieved by the RZFB (\ref{igs1}) and the new RZFB (\ref{migs1}).  Fig. \ref{RZF5} plots the achievable minimum IU throughput versus the number $M$ of BS antennas, clearly showing that Alg 3-IGS outperforms its counterpart Alg 3-PGS. Similarly,  Alg 4-IGS and Alg 5-IGS outperform their counterparts Alg 4-PGS and Alg 5-PGS.
 Fig. \ref{RZF6}, which plots the achievable minimum IU throughput for $K = M +1$, follow the same trend as Fig. \ref{RZF5}: Alg 3-IGS , Alg 4-IGS and Alg 5-IGS outperform their counterparts Alg 3-PGS, Alg 4-PGS and Alg 5-PGS, respectively. The advantage of the new RZFB over the conventional RZFB is also confirmed. Furthermore, Algorithm \ref{alg4} benefits to a lesser extent from the new RZFB than Algorithm \ref{alg3} and Algorithm \ref{alg5}.

Finally, Table \ref{table1} provides the average number of required iterations for the convex optimization part of the algorithms' convergence in
simulating  Fig. 9. The average single iteration time is 3.02s and 4.83s for the PGS based and IGS based algorithms, respectively.
All the algorithms only need $30\%$ of the maximum number of iterations to reach  $80\%$ of their optimal values.
\begin{table}[t]
\caption{The average number of iterations required for the algorithm's convergence}
\centering
\begin{tabular}{|l|c|c|c|c|c|}
\hline
 &$M=5$& $M=6$& $M=7$&$M=8$&$M=9$\\
\hline
Alg 3-PGS  &15&14&14&10&9\\
\hline
Alg 4-PGS & 10&9&8&8&8\\	
\hline
Alg 5-PGS & 17&16&17&14&11\\	
\hline
Alg 3-IGS &  17&16&17&14&13\\
\hline
Alg 4-IGS & 17&15&15&15&11\\
\hline
Alg 5-IGS &  23&21&23&17&12\\
\hline
\end{tabular}\label{table1}
\end{table}

\section{Conclusions}
We have considered a network in which a multi-antenna  aided BS and an RIS support multiple IUs and EUs. To facilitate computational tractability while aiming for the maximum possible information and energy throughput, conjugate beamforming has been used for delivering energy,  while zero-forcing  or regularized zero-forcing beamforming  has been used for delivering information under the transmit-TS framework, where energy and information are separately delivered during different time-slot fractions. The problem of jointly designing the RIS  PRCs and the power allocation of the beamformers for maximizing the minimum IU throughput subject to QoES in terms of the harvested energy thresholds at the EUs end has been addressed. It has been shown that this joint design can be decomposed into  separate designs of the RIS PRCs and of the power allocation of the IUs' beamforming. We have developed several efficient algorithms for these designs. A new regularized zero-forcing beamforming method has also been conceived for improving the IUs' throughput, which can improve the IUs' throughput significantly, especially in the regime of low numbers of BS antennas.
\section*{Appendix: rate function approximation}
The following inequality follows from the fact that the function $\la [\bV]^2\bY^{-1}\ra$ is convex for the the matrix
variable $\bV$ and positive matrix variable $\bY$ \cite{RTKN14}:
\begin{equation}\label{fund1}
\la [\bV]^2\bY^{-1}\ra \geq 2\Re\{\la \bar{V}^H\bar{Y}^{-1}\bV\}-\la [\bar{V}]^2\bar{Y}^{-1}\bY\bar{Y}^{-1}\ra,
\end{equation}
for all $\bV$, $\bar{V}$, and positive definite $\bY$ and $\bar{Y}$ of an appropriate dimension.

The following inequalities were obtained in \cite{TTN16}:
\begin{eqnarray}\label{fund4}
\ln\left(1+\frac{\bv^2}{\by}\right)&\geq&\ln\left(1+\frac{\bar{v}^2}{\bar{y}}\right)-
\frac{\bar{v}^2}{\bar{y}}  +2\frac{\bar{v}\bv}{\bar{y}} \nonumber\\
&&-\frac{\bar{v}^2\left(\by+\bv^2\right)}{\bar{y}(\bar{y}+\bar{v}^2)},
\end{eqnarray}
for all $\bv\in\mathbb{R}, \by>0$ and $\bar{v}\in\mathbb{R}, \bar{y}>0$, and
\begin{eqnarray}
\ln|I_{2}+[\bV]^2\bY^{-1}|&\geq& \ln|I_{2}+[\bar{V}]^2\bar{Y}^{-1}|-\la[\bar{V}]^2\bar{Y}^{-1}\ra\nonumber\\
&&-\la \bar{Y}^{-1}-\left([\bar{V}]^2+\bar{Y}\right)^{-1}, [\bV]^2+\bY\ra\nonumber \\
&& +2\Re\{\la \bar{V}^H\bar{Y}^{-1}\bV\ra\},\label{fund5}
\end{eqnarray}
for all matrices $\bV$ and $\bar{V}$, and positive definite matrices $\bY$ and $\bar{Y}$ of size $2\times 2$.

Particularly,
\begin{eqnarray}\label{fund6}
\ln\left(1+\frac{\bv^2}{\sigma}\right)&\geq&\ln\left(1+\frac{\bar{v}^2}{\sigma}\right)
-\frac{\bar{v}^2}{\sigma}  +2\frac{\bar{v}\bv}{\sigma}\nonumber\\
&&-\frac{\bar{v}^2\left(\sigma+\bv^2\right)}{\sigma(\sigma+\bar{v}^2)}
\end{eqnarray}
for $\sigma>0$ and $\bv\in\mathbb{R}$, $\bar{v}\in\mathbb{R}$.
\bibliographystyle{ieeetr}
\bibliography{surface}

\begin{IEEEbiography}[{\includegraphics*[width=1in,height=1.25in,clip,keepaspectratio]{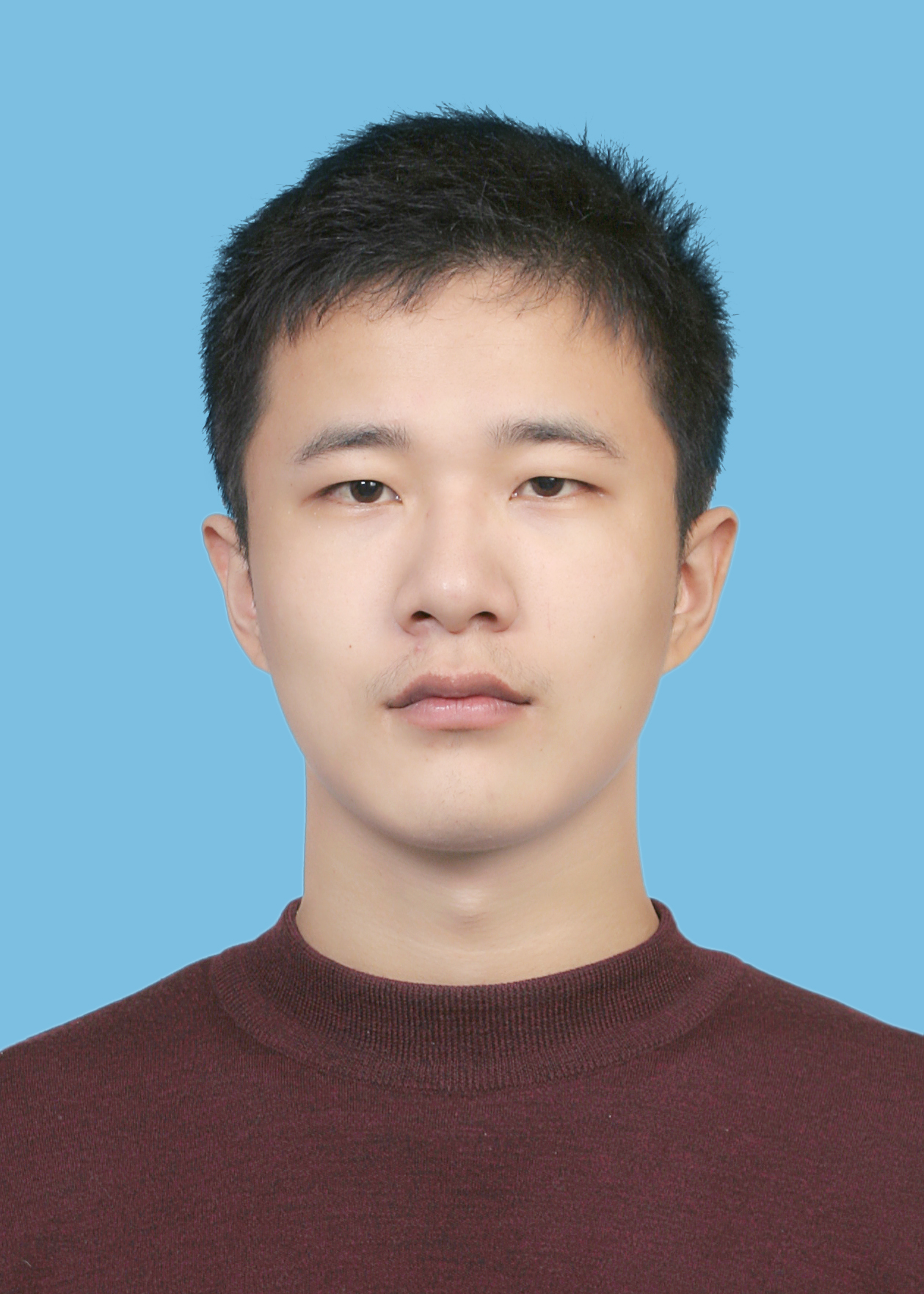}}]
{Hongwen Yu} received the Bachelor and Master degree from Shanghai University, Shanghai, China in 2011 and 2014, respectively, and received Ph.D. degree in Communication and Information Engineering from the Shanghai University, Shanghai, China, in 2020. He is currently pursuing the Ph.D. degree with the School of Electrical and Data Engineering, University of Technology Sydney, Ultimo, NSW, Australia. His current research interests include optimization methods for wireless communication and signal processing.
\end{IEEEbiography}

\vspace{-1cm}
\begin{IEEEbiography}[{\includegraphics*[width=1in,height=1.25in,clip,keepaspectratio]{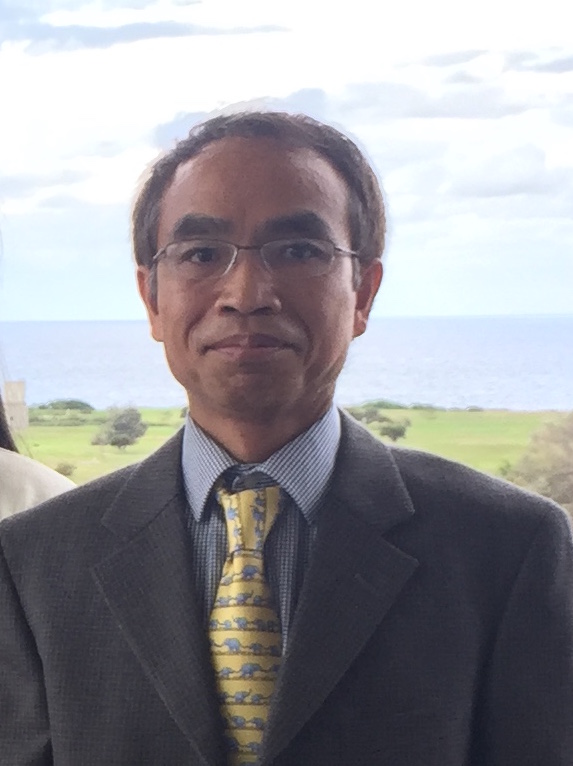}}]
{Hoang Duong Tuan} received the Diploma (Hons.) and Ph.D. degrees in applied mathematics from Odessa State University, Ukraine, in 1987 and 1991, respectively. He spent nine academic years in Japan as an Assistant Professor in the Department of Electronic-Mechanical Engineering, Nagoya University, from 1994 to 1999, and then as an Associate Professor in the Department of Electrical and Computer Engineering, Toyota Technological Institute, Nagoya, from 1999 to 2003. He was a Professor with the School of Electrical Engineering and Telecommunications, University of New South Wales, from 2003 to 2011. He is currently a Professor with the School of Electrical and Data Engineering, University of Technology Sydney. He has been involved in research with the areas of optimization, control, signal processing, wireless communication, and biomedical engineering for more than 20 years.
\end{IEEEbiography}

\vspace{-1cm}
\begin{IEEEbiography}[{\includegraphics*[width=1in, height=1.25in, clip, keepaspectratio]{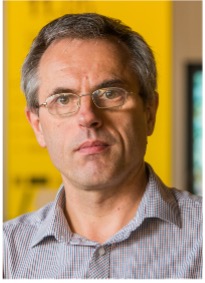}}]
{Eryk Dutkiewicz} received his B.E. degree in Electrical and Electronic Engineering from the University of Adelaide in 1988, his M.Sc. degree in Applied Mathematics from the University of Adelaide in 1992 and his PhD in Telecommunications from the University of Wollongong in 1996. His industry experience includes management of the  Wireless Research Laboratory at Motorola in early 2000 ’s. Prof. Dutkiewicz is currently the Head of School of Electrical and Data Engineering at the University of Technology Sydney, Australia. He is a Senior Member of IEEE. He also holds a professorial appointment at Hokkaido University in Japan. His current research interests cover 5G/6G and IoT networks.
\end{IEEEbiography}

\vspace*{-1cm}
\begin{IEEEbiography}[{\includegraphics*[width=1in,height=1.25in,clip,keepaspectratio]{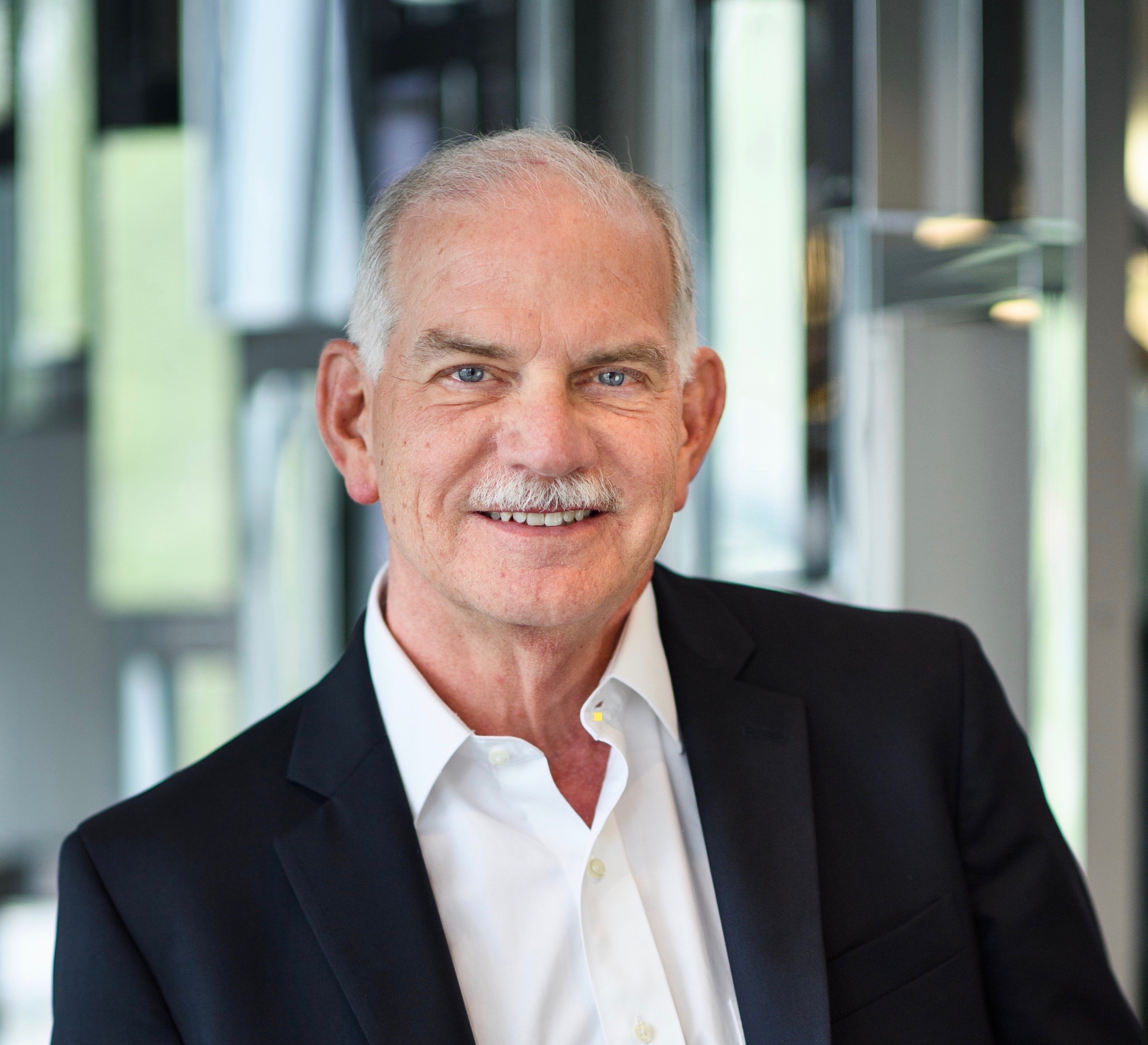}}]
{H. Vincent Poor} (S72, M77, SM82, F87) received the Ph.D. degree in EECS from Princeton University in 1977.  From 1977 until 1990, he was on the faculty of the University of Illinois at Urbana-Champaign. Since 1990 he has been on the faculty at Princeton, where he is currently the Michael Henry Strater University Professor. During 2006 to 2016, he served as the dean of Princeton's School of Engineering and Applied Science. He has also held visiting appointments at several other universities, including most recently at Berkeley and Cambridge. His research interests are in the areas of information theory, machine learning and network science, and their applications in wireless networks, energy systems and related fields. Among his publications in these areas is the forthcoming book {\it Machine Learning and Wireless Communications}  (Cambridge University Press).
Dr. Poor is a member of the National Academy of Engineering and the National Academy of Sciences and is a foreign member of the Chinese Academy of Sciences, the Royal Society, and other national and international academies. He received the IEEE Alexander Graham Bell Medal in 2017.
\end{IEEEbiography}

\vspace*{-1cm}
\begin{IEEEbiography}[{\includegraphics*[width=1in, height=1.25in, clip, keepaspectratio]{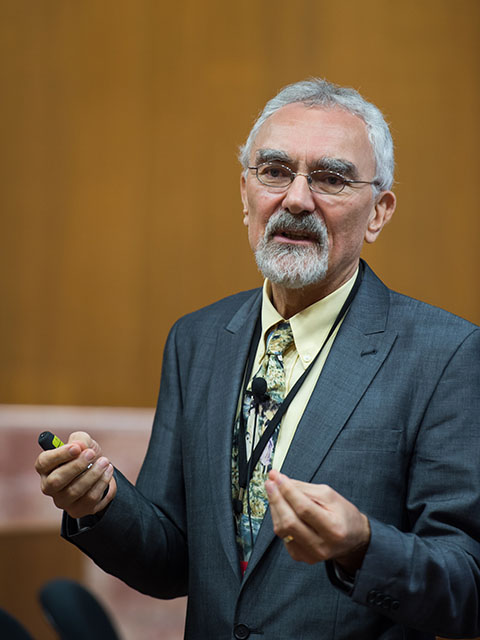}}]
{Lajos Hanzo}  (http://www-mobile.ecs.soton.ac.uk, https://en.wikipedia.org/wiki/Lajos$_{-}$Hanzo) (FIEEE'04) received his Master degree and Doctorate in 1976 and 1983, respectively from the Technical University (TU) of Budapest. He was also awarded the Doctor of Sciences (DSc) degree by the University of Southampton (2004) and Honorary Doctorates by the TU of Budapest (2009) and by the University of Edinburgh (2015).  He is a Foreign Member of the Hungarian Academy of Sciences and a former Editor-in-Chief of the IEEE Press.  He has served several terms as Governor of both IEEE ComSoc and of VTS.  He has published 2000+ contributions at IEEE Xplore, 19 Wiley-IEEE Press books and has helped the fast-track career of 123 PhD students. Over 40 of them are Professors at various stages of their careers in academia and many of them are leading scientists in the wireless industry. He is also a Fellow of the Royal Academy of Engineering (FREng), of the IET and of EURASIP.
\end{IEEEbiography}

\end{document}